\documentclass[a4paper,11pt]{article}
\pdfoutput=1
\usepackage[utf8]{inputenc}
\usepackage{jheppub}
\usepackage{amsfonts,amssymb, amsmath,url,hyperref,exercise,ulem,tensor,bm}
\hypersetup{
colorlinks = true,
linkcolor=blue,
anchorcolor=black,
citecolor=black,
filecolor=cyan,
menucolor=black,
runcolor=cyan,
urlcolor=magenta
}
\usepackage{mathtools}

\newcommand{\oao}[2]{{#1\atopwithdelims[]#2}} 
\def\sl2{SL(2,$\mathbb{R}$)}

\def\bE{\begin{Exercise}}
\def\eE{\end{Exercise}}
\def\bEL{\begin{ExerciseList}}
\def\eEL{\end{ExerciseList}}

\def\be{\begin{equation}}
\def\ee{\end{equation}}
\def\barr{\begin{array}{lr}}
\def\earr{\end{array}}
\def\bea{\begin{eqnarray}}
\def\eea{\end{eqnarray}}

\def\nn{\nonumber}

\def\del{\partial}

\def\S{{\cal S}}

\def\w{\omega}
\def\a{\alpha}
\def\b{\beta}

\def\d{\delta}
\def\e{\epsilon}

\def\f{\phi}

\def\k{\kappa}
\def\l{\lambda}
\def\m{\mu}
\def\n{\nu}

\def\p{\pi}

\def\r{\rho}
\def\s{\sigma}
\def\t{\tau}
\def\u{\upsilon}

\def\y{\psi}

\def\D{\Delta}

\def\L{\Lambda}
\def\O{\Omega}

\def\S{\Sigma}

\definecolor{maroon}{rgb}{0.5, 0.0, 0.0}	
\definecolor{arsenic}{rgb}{0.23, 0.27, 0.29}

\def\red#1{\textcolor{red}{#1}}
\def\blue#1{\textcolor{blue}{#1}}

\title{Revisiting the Thermal AdS partition function}
\author[1]{Roshan Kaundinya,}
\author[2]{Omkar Nippanikar,}
\author[3]{Akash Singh,}
\author[4]{K. P. Yogendran}
\affiliation[1]{Institute for Mechanical Systems, ETH Z\"urich, 8092 Z\"urich, Switzerland}
\affiliation[2]{Department of Theoretical Physics, TIFR, Mumbai}
\affiliation[3,4]{Department of Physics, IISER Mohali, \\ 
Sector 81, Knowledge City, Mohali, Punjab, 140306, India}
\emailAdd{roshan.kaundinya@gmail.com}
\emailAdd{omkar.nippanikar@gmail.com}
\emailAdd{akashsingh@iisermohali.ac.in}
\emailAdd{yogendran@iisermohali.ac.in}

\abstract{We rewrite the worldsheet torus partition function of the Thermal AdS CFT by isolating the boundary parameters. Using this, we show that the spectrum of the Euclidean BTZ black hole and Lorentzian AdS3 can be extracted -- the latter as a zero temperature limit. A similar procedure recovers the Lorentzian BTZ spectrum proposed in an earlier work. We then use our expression to construct a boundary modular invariant expression as a Poincar\'e series.}
\arxivnumber{}

\begin{document}

\maketitle
In this letter, we discuss the CFT worldsheet partition functions of global $AdS_3$,  the Lorentzian and Euclidean BTZ black holes, and Thermal $AdS_3.$ These geometries are closely related to each other giving us several reasons to compare them. While a modular invariant torus partition function for Thermal $AdS_3$ (TADS) was constructed in \cite{MO2}, that of Lorentzian $AdS_3$ (LADS) was proposed in \cite{Israel}. Even though the relation between the Euclidean BTZ (EBTZ) and TADS as orbifold CFTs was known from the work of \cite{Malstrom}, there remained a question of how to reinterpret the expression in \cite{MO2} in EBTZ terms. 
A somewhat basic question is the relation between the spectra of the Euclidean and Lorentzian spacetimes. 
The authors of \cite{MO2} interpreted their result as a bulk {\it free energy} to obtain the spectrum of states of the Lorentzian $AdS_3$ CFT proposed in \cite{MO1}. A question of interest therefore is to ask if the spectrum of Lorentzian $AdS_3$ can be obtained in a more direct manner as a zero temperature limit. 

The conformal boundary of these Euclidean spacetimes is a torus with a complex structure determined by the temperature and chemical potential. The worldsheet partition functions, after including ghosts and integrating over the worldsheet modular parameter will be a function of the boundary modular parameter. In string theory, the exponentiated torus amplitude gives a one loop contribution to the spacetime partition function which, by
the logic of the AdS/CFT correspondence, is interpreted as the partition function of the dual CFT residing on the boundary torus. Hence, it should be invariant under modular transformations acting on the boundary torus modular parameter.

Perhaps, the most important reason to study this system with more care is that it is a finite temperature black hole spacetime whose various avatars can all be exactly solved (as perturbative string theories) in principle. Since such systems are at a premium, a careful ground work can set the stage for many future studies. 

In this work, we rewrite the partition function of \cite{MO2} in a manner that aids the exploration of these questions. This rewriting  allows us to expand the partition function in a manner appropriate to the Euclidean BTZ black hole thereby identifying its spectrum. This expansion also points out that while the (Euclidean) NS-NS B-field vanishes at $r=0$ in Thermal AdS, it vanishes at the horizon in the Euclidean BTZ black hole. We show that this translates into the statement that the operator dual to the bulk B-field has different VEVs in these two states. 

We then show how the spectrum of Lorentzian $AdS_3$ may be extracted as a zero temperature limit. In fact, the entire partition function matches the proposal presented in \cite{Israel}. We also show how to Wick rotate the Euclidean BTZ spectrum into the Lorentzian black hole and reproduce the results in \cite{Nippanikar}. In this case though, the limit is not a zero temperature limit. 
Finally, using further Poisson resummations, we present a Poincar\'e series for the partition function of the boundary CFT together with a sketch of how the Hawking-Page transition might appear. 


We conclude with a summary and discuss some future questions for study.

\section{The Euclidean Geometries}
In this section, we will discuss Thermal AdS and the Euclidean BTZ black holes from the vantage point of the AdS/CFT correspondence. The analysis in this section is essentially a repeat of earlier work \cite{DMMV}, but we will obtain a couple of additional insights. Regarding these as solutions of a gravitational theory will allow us to characterise the difference between these geometries in terms applicable to a dual boundary CFT.

The bulk string theory can be viewed in three different lights: as Einstein gravity with a cosmological constant, as a 3D string theory with a $B_{NS}$ field, or as a 10-D string theory again with $B_{NS}$ fields. In the second case, these geometries are to be thought of as arising from a compactification of string theory. In the last case, one views the 3D-geometry as a factor in a product geometry defining a 10-D spacetime such as, for instance, the stringy S-dual of the D1-D5 system. 

In any case, the observation that Thermal AdS and Euclidean BTZ are both orbifolds of the 3D upper half plane $\mathbb{H}^3$ affords us a solvable worldsheet description of the 3D string theory. The orbifolding $\mathbb{Z}$ subgroup defines two real numbers which translate into the temperature and an angular chemical potential of the dual field theory. 

For setting up the AdS/CFT perspective, we will view the bulk geometries as solutions of a (super)gravity system whose action can be taken to be
\be
S=\frac1{2\k_3 ^2}\int d^3x\  e^{-2\Phi}\sqrt{g}\left(R-2\L+4(\nabla\Phi)^2-\frac12|H|^2\right).
\ee 
The cosmological constant is necessary to interpret the solution as the 3D description of the F1-NS5 system in string theory. 
The pure gravity viewpoint will include only a negative cosmological constant. In either case, we are to seek Euclidean solutions that are asymptotically AdS and whose boundary has a torus topology with periodicities defining the temperature and chemical potential of the dual CFT state. 


A torus is defined as a quotient of the complex plane with metric $ds^2=\r_2 dz d\bar{z}$,
by a lattice: 
\be 
z\sim z+2\pi \qquad z\sim z+2\pi T.
\ee
We can change coordinates  $z=2\pi(\phi'_E+ T t'_E)$ with $T=T_1+iT_2$ so that the metric takes the form:
\be
g_{MN}=\frac{\r_2}{T_2} \begin{pmatrix} 1 & T_1\\T_1& |T|^2\end{pmatrix} \label{torusmetric}
\ee
with $\r_2=(2\p)^2$ and the identifications become
\be
\label{undone}
(\phi'_E,t'_E)\sim(\phi'_E+1,t'_E)\,,\qquad (\phi'_E,t'_E)\sim(\phi'_E,t'_E+1)\,.
\ee
In the metric above,  $\r_2$ has dimensions $L^2$ while $T$ is dimensionless and the  coordinate fields $X^\m(\t,\s)=(\phi'_E,t'_E)$ of the string sigma model are dimensionless.
However, two parameters $T$ and $\frac{aT+b}{cT+d}$ with $a,b,c,d$ integers satisfying $ad-bc=1$ define the same torus.

When comparing two Euclidean geometries which are being viewed as bulk saddle points defining AdS duals corresponding to the same {\it state} of the boundary theory, we will only need that the asymptotic torii be the same upto such modular transformations. Thus, before comparing two such geometries as candidate descriptions of a single state of the boundary theory, we should first use suitable modular to transformations to bring $T$ to within the fundamental domain of the Teichmuller space $\mathcal{H}/PSL(2,Z).$ This fundamental domain parametrizes inequivalent torii.  Further, the volume $\r_2$ of the boundary torii needs to be the same. In addition, for Euclidean spacetimes, there is the question of choosing the time direction. 

In this section, we will manage these problems by rewriting the Thermal AdS and Euclidean BTZ geometries so that the boundary torus takes the form given in \eqref{torusmetric}. Then, we will simply require that the non-normalizable modes of all components of the bulk metric be equal (at a particular UV cutoff).

One solution with an asymptotic boundary torus is the Thermal $AdS_3$ (TADS) geometry with metric: 
\be
ds^2= \frac{l^2}{r^2+l^2}dr^2+\frac{r^2+l^2}{l^2} dt^2+\frac{r^2}{l^2} dx^2 \label{TADSm}
\ee
where $x\sim x+2\pi R$ with R being the radius of the boundary circle. All coordinates are dimensionful. If $l\neq R$, we have a conical singularity at $r=0$. We will use the parameter $R$ as a book-keeping device throughout.



The thermal identification
\be
(t,x) \sim (t+\b, x+\tilde\m\b)
\ee 
defines the inverse temperature $\b$ and a dimensionless chemical potential $\tilde\m$. 



We will rewrite this metric so that the identifications become as in \eqref{undone}. Using $R\f_E=x-\tilde\m t$ and $(\phi_E',t'_E)\coloneqq(\frac{\phi_E}{2\pi},\frac{t}{\b})$, this leads to:
\bea
ds^2_E=&\frac{l^2 }{r^2+l^2}dr^2  + \frac{(2\pi R)^2}{l^2}\left[r^2 d{\phi'_E}^2 + 2r^2\tilde\m\tilde\b d\phi'_E dt'_E + \tilde\b^2\left(l^2 + (\tilde\m^2+1)r^2\right) d{t'_E}^2\right]\label{TADSmetric}
\eea
with a dimensionless temperature $\tilde \b=\frac{\b}{2\p R}$ determined in terms of the boundary radius $R$. 
The important observation is the appearance of a dimensionless volume 
parameter $V=\frac{2\pi R \b}{l^2}$ of the boundary torus quite apart from the temperature and chemical potential. For later use in constructing the partition function, note that the boundary coordinates can be made dimensionful $(l\f'_E,lt'_E)$ by rescaling with the AdS radius. 

In case we regard the above as part of a string theory compactification, we will also have an imaginary $B_{NS}$ field with dimensions $L^{2}$ (which arises from the WZW term in the string sigma model): 
\be 
B_{NS}=\frac{i}{l^2}(r^2-C) dt\wedge dx= i\frac{(2\p R)^2\tilde\b}{l^2}(r^2-C) d\f' _E\wedge dt' _E
\ee
from which we get $H=2i\sqrt{g}$ which solves the gravity equations. 
Since the $x$ circle shrinks to zero size at $r=0$ making the coordinate ill-defined, $C=0$ for the two form to be sensible. 



Another space with a torus boundary is the Euclidean BTZ black hole (EBTZ) usually presented as the metric:
\be
ds^2_E =\frac{1}{l^2}\left[(r^2-r_+ ^2+r_- ^2)dt ^2-2r_+ r_-  dt\ dx+r^2dx^2\right]+\frac{l^2r^2dr^2}{(r^2-r^2_+)(r^2+r^2_-)}\label{BTZmetric}
\ee
where $x\sim x+2\pi R.$ 

At the horizon $r=r_+$, the $t$ coordinate shrinks to zero size defining a plane together with the radial coordinate $r$. This forces a periodicity on the $t$ coordinate $t\sim t_E+\b_H$ with the temperature $\b_H=\frac{2\pi r_+l^2}{(r_+ ^2+r_- ^2)}$ from the requirement that the geometry {\it terminate} smoothly at $r=r_+$. Simultaneously, the $x$ coordinate is assigned a period $x\sim x+\tilde\m\b_H$ with $\tilde\m=\tilde\m_H=\frac{r_-}{r_+}$. 

By defining $x=R\f_E+\tilde\m t$ and rescaling $(\phi_E',t'_E)\coloneqq(\frac{\phi_E}{2\pi},\frac{t}{\beta_H})$, we ensure that the identifications are as in \eqref{undone}
while the metric becomes
\begin{multline}
ds^2_E = \frac{(2\p R)^2}{l^2}\biggl[{\tilde\b_H}^2\left(r^2(1+\tilde\m^2)-r_+ ^2+r^2_- -2\tilde\m r_+r_-\right)d{t'_E}^2+ r^2 d{\phi'_E}^2 \\
+ 2\tilde\b_H(\tilde\m r^2-r_+r_-)dt'_E d\phi'_E \biggl]+ \frac{r^2dr^2}{(r^2-r^2_+)(r^2+r^2_-)}\label{EBTZmetric}
\end{multline}
where $\tilde\b_H=\frac{r_+ l^2}{R(r_+ ^2+r_-^2)}.$
Again, we will have a Euclidean $B_{NS}$ field which gives $H=2i\sqrt{g}=dB$:
\be 
B_E= i \frac{(2\p R^2)\tilde \b_H }{l^2}(r^2 - \bar{\a}) d\phi'_E \wedge dt'_E
\ee
where $\bar{\a}$ is an integration constant. Since the $t_E$ co-ordinate becomes ill-defined at $r=r_+$, the two form $B_E$ must vanish at $r=r_+$ which forces $\bar{\a}=r_+^2.$ 

This is in contrast with the previous geometry where it is the $x$ circle which shrinks to zero size and that too at $r=0$. From the string theory point of view, we can regard this difference between the two geometries as being captured by the B-field. The B-fields in the two cases differ by a constant, but this constant cannot be changed by a ``small" gauge transformation \cite{sashok}. As we will see below, the two geometries also differ in the expectation value of the operator dual to the B-field.

The boundary data can be obtained from the bulk fields by using the Fefferman-Graham expansion. This is achieved by redefining the radial coordinate so that the bulk metric takes the form of a series
\be 
ds^2_g = \frac{l^2d\rho^2}{\rho^2} +  \frac{l^2}{\rho^2} (G_{MN})_g dX^M  dX^N + l^2 (\tilde T_{MN})_g dX^M dX^N + \frac{\rho^2}{l^2} (F_{MN})_g dX^M dX^N + \ldots
\ee 
where $X^1 = \phi'_E$ and $X^2 = t'_E$ are dimensionless boundary coordinates and the subscript $g$ labels the geometries. The tensor  $G_{MN}$ is the induced metric on the conformal boundary and $T_{MN}$ with dimensions $L^{0}$ is interpreted as the (VEV of the) boundary energy-momentum operator.
\be 
T_{MN}=\frac{2 l}{16\p G_3} \tilde T_{MN}
\ee 
For thermal AdS \eqref{TADSm}, we get relation 
\be
\log\frac{\r}{\r_A}=\log\left(\frac{r+\sqrt{r^2+l^2}}{2l}\right)
\ee
containing a integration constant $\r_A$ with length dimensions. 
This gives us the {\it boundary} metric of Thermal AdS:
 \be \label{Ametrics}
(G_{MN})_{TAdS_3} =  \frac{\r_A ^2(2\pi R)^2}{16l^2} \begin{pmatrix} 
1 & \tilde\m\tilde\b \\
\tilde\m\tilde\b & \tilde\b^2(1+{\tilde\m}^2)
\end{pmatrix}
\ee 
which is dimensionful because the coordinates are dimensionless. 
A similar computation for the BTZ geometry gives: 
\be
\log\frac{\r}{\r_B}=\log\frac{\sqrt{r^2+r_-^2}+\sqrt{r^2-r_+^2}}{2\sqrt{r_+^2+r_-^2}}
\ee
contains the integration constant $\r_B$. Using this, the boundary metric is obtained to be: 
\begin{align} \label{Bmetrics}
(G_{MN})_{EBTZ} = \frac{\r_B ^2(r_+ ^2+r_- ^2)(2\pi R)^2}{16l^4} \begin{pmatrix} 
1 & \tilde\m \tilde\b_H \\
\tilde\m\tilde\b_H & \tilde\b_H ^2(1+{\tilde \m} ^2)
\end{pmatrix}
\end{align}

We can compare the metrics  above with the torus metric \eqref{torusmetric} and read off the modular parameter. The modular parameter for thermal AdS turns out to be 
$T_A=\tilde\b(i+\tilde\m)$ and  $T_B=\tilde\b_H(i+\tilde\m_H)$ for the black hole. Note that the two geometries have the same chemical potential and temperature if $\b=\b_H$ and $\tilde \m=\tilde\m_H.$ Using $\tilde\b_H=\frac{r_+l^2}{R(r_+^2+r_-^2)}$, and $\tilde\m=\m_H\equiv\frac{r_-}{r_+}$:
\be
T_B=\tilde\b_H(i+\tilde\m_H)=-\frac{l^2}{R(-r_-+ir_+)}\label{TempModBTZ}.
\ee
i.e., an S-modular transformation of a torus with modular parameter $(-r_-+ir_+).$ Thus, one could have expected that by substituting $\b(i+\m)=\frac{-l^2}{R(-r_-+ir_+)}$ in the Thermal AdS partition function  \cite{RR}, we should obtain the BTZ partition function. However, this naive expectation needs an important modification because the boundary torii have different volumes in the two cases.
To see this, we first read off the boundary volume parameters by comparison: 
\be
(\r_2)_{TADS}=\frac{\r_A ^2 R^2}{4l^2}\tilde\b \pi^2=\frac{\pi\r_A ^2\b R}{8l^2}  \qquad  (\r_2)_{EBTZ}=\frac{\r_B ^2 R^2 \p^2(r_+ ^2+r_-^2)\tilde\b_H}{4l^4}= \frac{\p^2 \r_B ^2 R r_+}{4l^2}\label{KahlerModuli}
\ee
for TAdS and EBTZ respectively. Note that the volume parameter $\r_2$ depends on size of the boundary circle $R$ and has dimensions of volume (because the coordinates are dimensionless). 

The integration constants can be fixed by requiring that $\lim_{r\to\infty} \frac{\r}{r}=1$ for both geometries.  
This gives, for TADS, $\r_A=l$ and the volume of the AdS boundary becomes 
\be
(\r_2)_{TADS}=\frac{\b R \pi}{8}
\ee
while for EBTZ, $\r_B=\sqrt{r_+^2+r_-^2}$ and the boundary torus has a different volume 
\be
(\r_2)_{EBTZ}=\frac{R\p^2(r_+ ^2+r_- ^2)r_+}{4 l^2}=\frac{R\p}{8|T|^2} \frac{\b}{|T|^2}.
\ee
rescaled by $|T|^4$. 
This observation needs to be taken into account in attempting to extract the BTZ spectrum from the Thermal AdS partition function of \cite{MO2}.

Alternatively, we have the option of choosing $\r_{A,B}$ to rescale the boundary volume to any value, for instance, $\r_2.$ This will allow ensure that the boundary torii are identical which then permits us to compare the on-shell actions and identify a Hawking-Page phase transition if any.

The boundary EBTZ metric can also be written as 
\be
\frac{(G_{MN})_{EBTZ}}{(2\p R)^2}=\frac{\r_B ^2}{16l^4 }(r_+ ^2+r_- ^2)\tilde\b_H ^2 (1+\tilde\m^2)\begin{pmatrix}
\frac{1}{\b_H^2(1+\m^2)} & \frac{\m}{\b_H(1+\m^2)} \\
\frac{\m}{\b_H(1+\m^2)}& 1
\end{pmatrix}.
\ee
Because 1 is in the ``wrong" position in comparison to the torus metric of \eqref{torusmetric}, we can read off the modular parameter only if we {\it rotate} $\f_E,t_E$:
\be
(G_{MN})_{EBTZ}\propto
(i\s_2)\begin{pmatrix}
1 & \frac{2\p\m}{\b_H(1+\m^2)} \\
\frac{2\p\m}{\b_H(1+\m^2)}& \frac{4\p^2}{\b_H^2(1+\m^2)}
\end{pmatrix}(-i\s_2)
\ee
which gives $T_B=\frac{-1}{\b_H(i+\m)}$ (we have omitted the prefactor in writing the above rotation).  
The rotation is the manifestation of the fact that the circle that shrinks to zero size is different in the two geometries. The form of the rotation matrix used in $G_{MN}$ implies that $(\f_E,t_E)\to (\f_E,t_E) .(i\s_2)=(t_E,-\f_E)$ or $z\to -iz.$

The rotation means that the identifications in the {\it original} coordinates are  
\be
\f_E+it_E\sim \f_E+it_E + \frac{2\p i }{\b_H(i+\m_H)}\sim \f+it_E - i(r_--ir_+) . \label{NewBTZIdent}
\ee
where the presence of $i$ signals the rotation 
This is the relation between the orbifold parameters of EBTZ and TADS noted in \cite{Malstrom}. 
In this version, we note that the volume parameters of TADS and EBTZ are equal if $\r_A ^2=\r_B^2.$ It now follows that, if we {\it replace} the Thermal AdS parameters $\b(i+\m) \to i(r_--ir_+)$, we will obtain the partition function of the Euclidean BTZ black hole provided we account for the rotation in interpreting the answer. We will use this approach in discussing the EBTZ torus partition function in section \ref{TP}.

Note that had we {\it interchanged} $x,y$ of the torus, the volume form in the bulk geometry would have changed sign - which at least in string theory, will mean that the B-field changes sign whereas, under {\it rotation} there is no sign change. This is important in order to ensure that the non-normalizable modes of the $B_{NS}$ field is the same for TADS and EBTZ geometries before we can compare them.

The normalizable modes of the bulk metric gives us the dimensionless stress-energy tensor of the boundary theory:
\begin{align} 
(T_{MN})_{EBTZ}= \begin{pmatrix} 
\frac{\p R^2 (r_+^2-r_-^2)}{4 G_3 l^3}& -\frac{ \p R r_-}{4 G_3 l}\\
\frac{\p R r_-}{4 G_3 l}&  -\frac{l \p}{4 G_3}
\end{pmatrix},
&&(T_{MN})_{TAdS_3}= \begin{pmatrix}
-\frac{\p R^2}{4 G_3 l}& -\frac{\tilde\m\b R }{8 G_3 l} \\
 -\frac{\tilde\m\b R }{8 G_3 l} &   \frac{\b^2(1-\tilde\m^2)}{16\p G_3 l} 
\end{pmatrix} \label{BStress}
\end{align}
  It should be noted that these are independent of the integration constant $\r_{A,B} ^2$ reflecting scale invariance as expected of a conformal field theory. The tracelessness of these stress-energy tensors can be verified using the boundary metric. 

\subsection{Thermodynamics and Phase transition} \label{HP}

If we wish to regard these two geometries as bulk saddle points corresponding to the same boundary state, we need to ensure that the boundary metrics are the same. Since $t_E$ and $\f_E$ have the same periodicities, we just need to compare the TADS metric in \eqref{Ametrics} with the first form given in \eqref{Bmetrics}, we see that the two are identical but for the prefactors. To match these, we need to set
the Kahler moduli $\r_A$ in terms of $\r_B$ as
\be
\r_A^2=\r_B ^2 M=\r_B ^2 (r_+ ^2+r_-^2).\label{kahler}
\ee
We also have to ensure that the non-normalizable modes of the B-field are equal. If we regard these two geometries as solutions to pure 3D gravity, this comparison is moot (since there is no $B_{NS}$ field in pure gravity). Therefore, in the {\it string theory} perspective, we expect this to be automatically satisfied provided the boundary metrics are the same. Applying the F-G expansion to the Euclidean B-field:
\be 
B_{EBTZ} = i (2\pi R)^2 \frac{\tilde\b_H}{l^2} \left(\frac{(r^2_++r^2_-)\r_B ^2 }{16\rho^2} - \frac{r^2_++r^2_-}{2} + \ldots \right)  d\phi'_E \wedge dt'_E.
\ee 
and for AdS, we have the expansion
\be 
B_{TAdS_3} = i (2\pi R)^2\frac{\tilde \b}{l^2} \left(\frac{l^2\r_A ^2}{16\rho^2} - \frac{l^2}{2}+ \ldots \right) d\phi'_E \wedge dt'_E.
\ee
The non-normalizable modes (sources) $J$  and normalizable modes (VEVs) $\langle \hat L\rangle $ are 
\begin{align}\nn
J_{EBTZ}&=& i(2\p R)^2\frac{r_+\r_B^2}{16}=i\frac{\p^2 \r_B^2}{4\b(1+\m^2)}
\qquad 
\langle \hat L\rangle_{EBTZ}&=&-i(2\p)^2r_+&& =-i\frac{1}{\b_H(1+\m_H ^2)}\\
J_{TAdS_3}&=& i(2\p R)^2\frac{\b \r_A^2}{4}
=i\frac{\pi \r_B^2 }{4\b(1+\m^2)} 
\qquad 
\langle \hat L\rangle_{TAdS}&=&-i\frac{\b(2\p)^2}{4}&&\label{B-fielddata}
\end{align}
where, in the second line, we have used the relation \eqref{kahler} and the relation between $r_+$ and the BTZ temperature to show that the sources $J$ are equal.  We also note that the sources are equal to the Kahler modulus $J=i\r_2$ \eqref{KahlerModuli}.

On the other hand, the expectation values of this operator are not equal implying that TAdS and EBTZ are different states of the boundary theory.  The operator $\hat{L}$ in the boundary theory which couples to the bulk $B-$field will have scaling dimension $\D=2.$ The relation between the non-normalizable mode and the normalizable mode 
$J=\frac{2\r_B ^2}{4\pi^2} \langle \hat{L}\rangle$
is an equation of state provided the operator $\hat{L}$ defines some kind of conserved ``charge" of the dual theory.


The geometries that we are studying should be regarded as defining a grand-canonical ensemble of a boundary theory because of the temperature periodicity and the angular periodicity which leads to a cross term in the metric. 
In such a situation, the phase of the theory is determined by the configuration corresponding to the lowest grand potential which we will determine by using the procedure of holographic renormalization.

We begin by regarding the two spacetimes as solutions of a gravity theory defined by 
\be
S=\frac{1}{16\pi G_3}\int d^3x \sqrt{g}(R-2\L).
\ee
where the cosmological constant $\L=-\frac{1}{l^2}.$ Even if we replace the cosmological constant with an NS-NS field $H$ or consider a 10-D bulk gravity action as the starting point, we obtain the same results.

For the Thermal AdS geometry, the Entropy ({\it defined} as the area of a horizon) is zero. The grand potential is evaluated by the procedure of Holographic Renormalization \cite{Skenderis} which requires the addition of specific counterterms to the gravitational action.  
\be 
G_{AdS}=\frac{S_{AdS}^E}{4\p^2 R \b}=\frac{s_{ads}+s_{GH}+s_{ct}}{4\p^2 R \b } 
\ee
Using these, 
we find that 
\be 
G_{AdS}=-\frac{1}{16 \p G_3 l}=-p
\ee
Note that since $G_{AdS}$ is independent of the temperature $\b$ and chemical potential $\m$, the entropy $s=-\frac{\del G}{\del T}$ and number density $n=-\frac{\del G}{\del \m}$ evaluate to zero. The Euler relation $\e+p=\m n+ sT$ implies that $\e=-p$ which agrees with $\e=\frac{\del(\b G)}{\del \b}$. 

For EBTZ on the other hand, the entropy evaluated by using the area law is:
\be 
S=\int \frac{d A}{4G_3}=\frac{2\p R}{4G_3 l}\int_{0}^{1}d\phi_E  r \big|_{r=r_{+}}=\frac{2\p R r_{+}}{4 G_3 l}. 
\ee
The Grand potential evaluated in the same manner as for Thermal AdS gives
\be 
G_{BH}=-\frac{1}{16 \p G_3 l^3}(r_{+}^2+r_{-}^2)
\ee 
We may determine the Entropy density from 
\be 
s=-\frac{\del G_{BH}}{\del T}\Big|_{\tilde\m}=\frac{ r_{+}}{4G_3 l}
\ee 
where $\tilde\m=\frac{r_{-}}{r_{+}}.$

The number density 
\be 
n=-\frac{\del G_{BH}}{\del \tilde\m}\Big|_{T}=-\frac{r_{+}r_{-}}{8\p G_3 l^3}
\ee 
and the pressure is 
\be
p=-G_{BH}=\frac{1}{16 \p G_3 l^3}(r_{+}^2+r_{-}^2)
\ee 
The Euler relation
\be 
\e+p=\tilde\m n +s T
\ee 
then gives us $\e=\frac{1}{16 \p G_3 l^3}(r_{+}^2-r_{-}^2)$ which agrees with the thermodynamic derivative.

Finally, we can discuss the Hawking-Page transition for this system. This involves comparing the pressures
\be
p_A=\frac{1}{16 \p G_3 l^3} \qquad p_B=\frac{1}{16 \p G_3 l^3}(r_{+}^2+r_{-}^2)
\ee
which tells us that when $r_+ ^2+r_-^2=\frac{l^2}{\beta^2(1+\m^2)}>1$, i.e., at high temperatures, the black hole, having the higher pressure, is the preferred description of the phase. This condition agrees with the condition obtained in \cite{Malstrom}, of course.


\section{Partition functions} \label{TP}
In this section, we will manipulate the CFT torus partition functions for these geometries.  We will first rewrite the torus partition function of Thermal AdS by isolating  a `torus' factor. This factor in the partition function will be modular invariant by itself and will contain all the dependence on the spacetime parameters $T$ and $\r$.  We will then expand these partition functions in q-series which will allow us to read off the spectrum. In particular, we read off the spectrum of the Euclidean BTZ black hole. We also show that the spectra agrees with semi-classical analysis and the orbifold interpretations. Finally, we will compare the partition functions themselves.

The partition function of thermal $AdS_3$ was computed in \cite{MO2}, by orbifolding a WZW model with target space $\mathbb{H}^3/SU(2).$ In the previous section, we introduced the TADS geometry which had a boundary circle of radius $R$ which defined the Kahler parameter $\r_2.$ 

This torus amplitude computed in \cite{MO2} is: 
\be 
Z_{TAdS_3}(\b,\m;\t) = \frac{\b \sqrt{k-2}}{2\pi \sqrt{\t_2}} \sum_{n,m\in\mathbb{Z}} \frac{e^{-\frac{k\b^2}{4\pi \t_2}|m-n\t|^2 + \frac{2\pi (\mathfrak{I}(U_{n,m}))^2}{\t_2}}}{|\vartheta_1(U_{n,m}|\t)|^2},\label{eqn:StartingPartFunc}
\ee 
where\footnote{Our $U_{n,m}$ is the complex conjugate of that used in~\cite{MO2} because we work in conventions where the left factor of the solution of the WZW model is holomorphic (left-moving on Lorentzian worldsheet). See Appendix~\ref{sec:AppendPathIntegral} for details.} $U_{n,m}\coloneqq\frac{\tilde\b}{2\p}(\tilde\m-i)(n\t-m)$ and $k$ is the level of the WZW model. We will henceforth omit the tilde in the temperature and chemical potential. 
Since the Kahler parameter $\r_2$ contains the boundary radius $R$, and in \cite{MO2}, the level $k$ sets the size of the boundary radius $l^2=k\a'$, the parameter $k$ in the above partition function is to be identified with $\r_2$. This fact will be important when we try to compare the partition functions of EBTZ and TADS.

\subsection{q-Expansion }

This partition function has only two integers, although we might expect to see four integers related to winding and momenta of the torus submanifold fibred over the radial direction. We will therefore apply some preliminary massaging. First rewrite the partition function using delta functions as
\begin{multline}
Z_{TAdS_3}(\b,\m;\t)=\frac{\b \sqrt{k-2}}{2\pi \sqrt{\t_2}} \sum_{n,m\in\mathbb{Z}} e^{-\frac{k\b^2}{4\pi \t_2}|m-n\t|^2}\\
\times\int_{-\infty}^{\infty} \mkern-18mudz_1 \int_{-\infty}^{\infty} \mkern-18mudz_2 \int_{-\infty}^{\infty} \mkern-18mudk_1 \int_{-\infty}^{\infty} \mkern-18mudk_2\;\frac{e^{\frac{2\pi z^2_2}{\t_2}} e^{2\p i\bigl[k_1(z_1-\mathfrak{R}(U_{n,m}))+k_2(z_2-\mathfrak{I}(U_{n,m}))\bigr]}}{|\vartheta_1(z_1+iz_2|\t)|^2}\,.
\end{multline}
Then, perform a Poisson Resummation in $m$ to obtain
\begin{multline}
Z_{TAdS_3}(\b,\m;\t)=\sqrt{\frac{k-2}{k}}\sum_{m^\prime,n\in\mathbb{Z}}\int d^2z\;d^2k\;\frac{e^{\frac{2\p}{\t_2}z_2^2}\;e^{2\p i\left[k_1\left(z_1-\frac{\b}{2\p}n\t_2\right)+k_2\left(z_2-\frac{\b}{2\p}\m n\t_2\right)\right]}}{|\vartheta_1(z_1+iz_2|\t)|^2}\\
\times\exp\left\{2\p i\t_1nm^\prime-\p\t_2\left[\frac{1}{k}\left(\frac{2\p}{\b}m^\prime-k_2+k_1\m\right)^2+k\left(\frac{\b n}{2\p}\right)^2\right]\right\} .
\end{multline}
We can now integrate over $k_2$ and perform a change of variables $z_2=v_1\t_2$ and $z_1=v_1\t_1-v_2$ which results in
\begin{multline}
Z_{TAdS_3}(\b,\m;\t)=\sqrt{\frac{k-2}{\t_2}}\sum_{m^\prime,n\in\mathbb{Z}}\t_2\int_{-\infty}^\infty\mkern-18mudv_1\int_{-\infty}^\infty\mkern-18mudv_2\int_{-\infty}^\infty\mkern-18mudk_1\\
\times\frac{e^{(2-k)\p v_1^2\t_2}\;e^{2\p i\left[k_1(v_1\t_1-v_2)+v_1\t_2\left(k_1\m+\frac{2\p}{\b}m^\prime-\frac{i\m k\b n}{2\p}\right)\right]}}{|\vartheta_1(v_1\t-v_2|\t)|^2}\\
\times\exp\left\{2\p i\t_1nm^\prime-\p\t_2\left[k\left(\frac{\b n}{2\p}\right)^2(1+\m^2)+2i\m m^\prime n+\frac{i\b nk_1}{\p}(1+\m^2)\right]\right\}\,.
\end{multline}%
%
%
%
%
We split $v_1=s_1+w_1$ and $v_2=s_2+w_2$, where $w_1,w_2\in\mathbb{Z}$ and $s_1,s_2\in[0,1)$. This results in 
\begin{multline}
Z_{TAdS_3}(\b,\m;\t)=\sqrt{\frac{k-2}{\t_2}}\sum_{\substack{m^\prime,n,\\w_1,w_2\in\mathbb{Z}}}\t_2\int_{0}^1\mkern-9muds_1 \int_0^1\mkern-9muds_2\int_{-\infty}^\infty\mkern-18mudk_1\;\frac{e^{2\p i\left[\t_1(nm^\prime+k_1(s_1+w_1))- k_1(w_2+s_2)\right]}}{|\vartheta_1(s_1\t-s_2|\t)|^2}\\
\times e^{-\p\t_2\left[-2i(s_1+w_1)\left(k_1\m+\frac{2\p}{\b}m^\prime-\frac{i\m k\b n}{2\p}\right)+ (k-2)s_1^2+k(2s_1w_1+w_1^2)+k\left(\frac{\b n}{2\p}\right)^2(1+\m^2)+2i\m m^\prime n+\frac{i\b nk_1}{\p}(1+\m^2)\right]}\,,
\end{multline}
upon using the following identity for the $\vartheta_1$ function
\be
{\frac{e^{(2-k)\p(s_1+w_1)^2\t_2}}{|\vartheta_1((s_1+w_1)\t-v_2|\t)|^2}=\frac{e^{(2-k)\p s_1^2\t_2}}{|\vartheta_1(s_1\t-v_2|\t)|^2}\;e^{-k\p\t_2(2s_1w_1+w_1^2)}}\,.
\ee
We can get rid of the integral in $k_1$ by replacing the sum $\sum_{w_2}e^{-2\pi i w_2 k_1}$ with $\sum_a \delta(k_1-a)$, resulting in 
\begin{multline}\label{ThePartFun}
Z_{TAdS_3}(\b,\m;\t)=\sqrt{\t_2(k-2)}\sum_{\substack{m^\prime,n,\\w_1,a\in\mathbb{Z}}} \int_0^1\mkern-9muds_1 \int_0^1\mkern-9muds_2\; 
\frac{e^{-2\p ias_2} e^{2\pi\t_2 s_1^2}}{|\vartheta_1(s_1\t-s_2|\t)|^2}\;e^{2\p i\t_1\left[nm^\prime+a(s_1+w_1)\right]}\\
\times e^{-\p\t_2\left[-2i(s_1+w_1)\left(a\m+\frac{2\p}{\b}m^\prime-\frac{i\m k\b n}{2\p}\right)+ k(s_1+w_1)^2+k\left(\frac{\b n}{2\p}\right)^2(1+\m^2)+2i\m m^\prime n+\frac{i\b na}{\p}(1+\m^2)\right]}\,.
\end{multline}
At this stage, it is worth pointing out that in the above partition function, we have isolated all dependence on the external parameters $\b,\m$ in the second line.

We will now expand the above expression following the ideas of \cite{MO2,Israel} so that we can compare the conformal weights with expectations based on classical analysis of strings moving on the Thermal AdS geometry. The details of this expansion are presented in Appendix~\ref{sec:AppendExpansion} and the result takes the form: 
\be
\begin{aligned}
Z(\b,\m;\t,\bar{\t})&=\sum_{m^\prime,n,w_1\in\mathbb{Z}}\sum_{q',\bar{q}',N,\bar{N}}\Biggl\{\int_{-\infty}^\infty\frac{ds}{\p}\left[\sum_{P_{\pm,0}=-\infty}^0\tfrac{1}{2is-kw_1+\left(i(q-\bar{q})\m+\frac{2\p i}{\b}m^\prime+\frac{\m k\b n}{2\p}\right)+1+q+\bar{q}}\right.\\
&\left.-\sum_{P_{\pm,0}=0}^\infty\tfrac{1}{2is-kw_1+\left(i(q-\bar{q})\m+\frac{2\p i}{\b}m^\prime+\frac{\m k\b n}{2\p}\right)-1+q+\bar{q}}\right]+\sum_{\substack{P_{\pm,0}=-\infty\\\frac{1}{2}+is=j_{q,\bar{q},w_1}^{n,m^\prime}\\-\frac{(k-2)}{2}<\mathfrak{I}(s)<0}}^0\Biggr\}\\
&\times e^{2\p i\t_1\left[w_1(q-\bar{q})+nm^\prime+N-\bar{N}\right]} \times e^{-2\p\t_2\left[\frac{2s^2}{k-2}-\frac{1}{4} +N+\bar{N}\right]}\\
&\times e^{-2\p\t_2\left[-w_1\left(i(q-\bar{q})\m+\frac{2\p i}{\b}m^\prime+\frac{\m k\b n}{2\p}\right)+\frac{k}{2}w_1^2+\frac{k}{2}\left(\frac{\b n}{2\p}\right)^2(\m^2+1)+i\m m^\prime n+\frac{i\b n(q-\bar{q})}{2\p}(1+\m^2)\right]}\,,
\end{aligned}\label{MOG2PartFuncChiralityFlip}
\ee
where the discrete series states have quadratic Casimir $j(1-j)$ with $j$ taking the values
\begin{equation}
j_{q,\bar{q},w_1}^{n,m^\prime}\coloneqq\frac{k}{2}w_1-\frac{1}{2}\left(i(q-\bar{q})\m+\frac{2\p i}{\b}m^\prime+\frac{\m k\b n}{2\p}\right)-\frac{1}{2}(q+\bar{q})\label{ThermalAdSDiscreteSeriesPoles}
\end{equation}
and $\sum_{q',\bar{q}',N,\bar{N}}$ is shorthand for a sum over infinitely many variables $P^+_{\pm,p},P^-_{\pm,p},P_{\pm,p}\in\{0,1,2,\ldots\}$ (for each $p\in\{1,2,3,\ldots\}$) with $(q,\bar{q})$ and $(N,\bar{N})$ defined as
\begin{subequations}
\begin{align}
q&\coloneqq P_{+,0}+q'\coloneqq P_{+,0}+\sum_{p=1}^{\infty}(P^+_{+,p}-P^-_{+,p})\,,\\
\bar{q}&\coloneqq P_{-,0}+\bar{q}'\coloneqq P_{-,0}+\sum_{p=1}^{\infty}(P^+_{-,p}-P^-_{-,p})\,,\label{qandbarq}\\
N&\coloneqq\sum_{p=1}^{\infty}p(P_{+,p}+P^+_{+,p}+P^-_{+,p})\,,\\
\bar{N}&\coloneqq\sum_{p=1}^{\infty}p(P_{-,p}+P^+_{-,p}+P^-_{-,p})\,.\label{NandbarN}
\end{align}
\end{subequations}
It is important to point out that the quantities in braces in \eqref{MOG2PartFuncChiralityFlip} are to be understood as a degeneracy of states with the specified values of $L_0, \bar{L}_0$. For the continuous series, the first two terms are combined into a density of states \cite{MO2}. The summations in the last term will give us the degeneracy of the states of the Discrete series which is clearly a positive integer.

There are two noteworthy features: the location of the poles involve $q+\bar q,$ but $q+\bar{q}$ does not appear in the continuous series $L_0+\bar{L}_0$. And, the discrete series states have both real and imaginary parts in the radial momentum $s.$ 
The imaginary parts will localize the wavefunctions in the interior of the geometry as expected since massive particles cannot reach the boundary of AdS.

\subsection{Interpretation -- thermal \texorpdfstring{$AdS_3$}{AdS3} viewpoint}\label{sec:TAdSViewpoint}
The action for the \sl2 WZW model at level $k$ written in the global coordinates $g=e^{\frac{i}{2}(t_{\text{global}}+\f)\s_2}e^{\r\s_3}e^{\frac{i}{2}(t_{\text{global}}-\f)\s_2}$ is
\begin{multline}
S_{\text{WZW}}[g]=\frac{-k}{2\p}\int dx^+dx^-\;\Big[\del_+\f\del_-\f\sinh^2\r+\del_+\r\del_-\rho-\del_+t_{\text{global}}\del_-t_{\text{global}}\cosh^2\rho\\
{+(\sinh^2\rho+C)}\left(\del_-t_{\text{global}}\del_+\phi-\del_-\phi\del_+t_{\text{global}}\right)\Big]\,,\label{WZWAction}
\end{multline}
where $C$ is a gauge ``ambiguity" in the B-field. Note that this action is manifestly Lorentzian. Thermal AdS is obtained from this by a Wick rotation and orbifolding. In thermal $AdS_3$, $\r=0$ is the origin of a polar coordinate system. The 1-form $d\f$ and hence, the 2-form $dt\wedge d\f$ are not defined at $\r=0$. So, it is expected that $C=0$ to ensure that the B-field vanishes at $\r=0$ where the geometry ``terminates".
The relation between these coordinates and those of the previous section are 
$r=\sinh\r$
and the AdS radius is related to $l^2=k\a'$ 

The Virasoro generators $L_0,\bar{L}_0$ for $AdS_3$ are constructed out of the currents of \sl2 WZW model.
The energy and angular momentum of TAdS are both obtained from the elliptic generator of \sl2. Thus, we can express \cite{MO1}
\begin{subequations}\label{L0pmL0barTAdS3}
\begin{align}
    L_0+\bar{L}_0&=\frac{1}{k}\biggl[(J^{(0)}_{+,0})^2-(J'^{(0)}_{+,0})^2+(J^{(0)}_{-,0})^2-(J'^{(0)}_{-,0})^2\biggl]+\frac{2s^2+\frac{1}{2}}{k-2}\,,\label{L0plusL0barTAdS3}\\
    L_0-\bar{L}_0&=\frac{1}{k}\biggl[(J^{(0)}_{+,0})^2-(J'^{(0)}_{+,0})^2-(J^{(0)}_{-,0})^2+(J'^{(0)}_{-,0})^2\biggl]\label{L0minusL0barTAdS3}\,,
\end{align}
\end{subequations}
where $J'^{(0)}_{\pm,0}$ denote eigenvalues of currents in the sector twisted by the orbifolding. The signs appearing in the above expression depend on the elliptic nature of the diagonalised generator. Wick rotation from Lorentzian AdS to Thermal AdS does not affect the elliptic nature of the generator, but it makes their eigenvalues complex.

We can verify that the choice
\begin{subequations}\label{CurrentsTAdS3}
\begin{align}
    J'^{(0)}_{\pm,0}&=\pm\frac{1}{2}\left[-\frac{2\p im^\prime}{\b}+(q-\bar{q})(-i\m\pm1)\right]-\frac{k}{2}\frac{i\b n}{2\p}\,,\\
    J^{(0)}_{\pm,0}&=J'^{(0)}_{\pm,0}\mp\frac{k}{2}\left(-w_1-\frac{\b n}{2\p}(-\m\pm i)\right)\,,
\end{align}
\end{subequations}
ensures that $L_0\pm\bar{L}_0$ read off from~\eqref{MOG2PartFuncChiralityFlip} satisfy~\eqref{L0pmL0barTAdS3}. 
We can now compare the currents with what is expected for thermal $AdS_3$ with inverse temperature $\b$ but chemical potential $-\mu$. 
The Noether charge corresponding to $\f$ and $t_{\text{global}}$ translations calculated from the action $S_{\text{WZW}}=\int d\t d\s\mathcal{L}$ are
\begin{subequations}
\begin{align}\label{TADSCons}
    \Pi_\f&\coloneqq-\int_0^{2\p}\mkern-18mud\s\;\frac{\del\mathcal{L}}{\del(\del_\t\f)}=J'^{(0)}_{+,0}+J'^{(0)}_{-,0}-\frac{ik\b n}{2\p}(C-1)=q-\bar{q}-kC\frac{i\b n}{2\p}\,,\\
    \Pi_{t_{\text{global}}}&\coloneqq-\int_0^{2\p}\mkern-18mud\s\;\frac{\del\mathcal{L}}{\del(\del_\t t_{\text{global}})}=J'^{(0)}_{+,0}-J'^{(0)}_{-,0}+kw_1C-kC\frac{\m\b n}{2\p}\\
    &=-\frac{2\p im'}{\b}-i\m(q-\bar{q})+kw_1C-kC\frac{\m\b n}{2\p}\,.
\end{align}
\end{subequations}
The target space momenta are correctly quantised for $C=0$ i.e.,
\begin{equation}
    \Pi_\f=q-\bar{q}\in\mathbb{Z}\,,\qquad\frac{i\b}{2\p}\Pi_{t_{\text{global}}}-\frac{\m\b}{2\p}\Pi_\f=m'\in\mathbb{Z}\,.
\end{equation} 
The relation between $J'^{(0)}_{\pm,0}$ and $J^{(0)}_{\pm,0}$ means that \eqref{CurrentsTAdS3} are the eigenvalues evaluated in the twisted sector with $(t_{\text{global}},\f)\xmapsto{\s\mapsto\s+2\p}(t_{\text{global}},\f)-n(i\b,-\m\b)-w_1(0,2\p)$. In particular, the integers $q-\bar{q}$ and $m'$ are (resp.) momenta along the $\f$ direction and the orbifold direction, while $w_1$ and $n$ are (resp.) the winding numbers along these directions. Note the presence of Matsubara terms in the energy 
\be
\Pi_{t_{\text{global}}}=\frac{2\p}{i\b} m'-i\m(q-\bar{q})
\ee
as appropriate to a thermal system.

Before moving on, a representation theoretic comment is in order. In the case of Lorentzian $AdS_3$, it is well known that highest and lowest weight discrete series with amounts of spectral flow differing by 1 and the radial momenta related by reflection about $\frac{k}{2}$ are identified. This also holds in thermal $AdS_3$ and is clear from the interpretation of the partition function discussed so far. To see this, note that
\be
    J^{(0)}_{\pm,0}=\pm(j+q_\pm)\,,\qquad J'^{(0)}_{\pm,0}=\pm(j+q_\pm)\pm\frac{k}{2}\O_\pm\,,
\ee
where we have defined $(q_+,q_-)\coloneqq(q,\bar{q})$, $\O_\pm\coloneqq-w_1+\frac{\b n}{2\p}(\m\mp i)$ and $j\coloneqq j^{n,m'}_{q,\bar{q},w_1}$ from~\eqref{ThermalAdSDiscreteSeriesPoles}, for convenience of notation. In this notation,~\eqref{L0pmL0barTAdS3} implies
\begin{subequations}\label{eqn:RepIdentify}
\begin{align}
    L_0&=\frac{-(j-\frac{1}{2})^2+\frac{1}{4}}{k-2}-\O_+(j+q)-\frac{k}{4}\O_+^2+N\\
    &=\frac{-(j'-\frac{1}{2})^2+\frac{1}{4}}{k-2}+\O'_+(j'-q)-\frac{k}{4}{\O'_+}^2+N+q
\end{align}
\end{subequations}
(and similarly for $\bar{L}_0$), where the radial momenta $j$ and $j'$ are related by a reflection about $\frac{k}{2}$ i.e., $j'=\frac{k}{2}-j$ and the spectral flow parameters are related by $\O'_\pm=\O_\pm+1$ i.e, by a shift of the $\f$ winding $w_1$ by 1. This shows that the lowest weight discrete series is identified with the highest weight discrete series with reflected radial momentum and shifted spectral flow, exactly as in Lorentzian $AdS_3$. This is a consequence of the interpretation~\eqref{CurrentsTAdS3} that states are labelled by eigenvalues of the elliptic generators $J'^{(0)}_{\pm,0}$. As we shall see, this identification will fail in the EBTZ intepretation of the same partition function where the states are labelled by eigenvalues of the hyperbolic generators $J'^{(2)}_{\pm,0}$ instead.

\subsection{Interpretation -- Euclidean BTZ viewpoint}\label{sec:EBTZViewpoint}
The Euclidean BTZ black hole is also an orbifold of the $\mathbb{H}^3$ in the same manner as Thermal AdS \cite{Malstrom}. Through~\eqref{NewBTZIdent}, we observed that the EBTZ can be viewed as an orbifold of $\mathbb{H}^3$ with a modular parameter $\tilde T_B=\pm (ir_-+r_+)$ provided we also rotate $t_E$ and $\f_E$ by $\frac{\p}{2}.$ 

In addition, we noted that using this second form of the boundary metric, the boundary volumes will be the same if $\r_A=\r_B.$ 
This means that by setting $(\b,\m\b)=(2\p r_+,2\p r_-)\in\mathbb{R}^2$, the same partition function~\eqref{ThePartFun} may be interpreted as the Euclidean BTZ partition function. 

Thus, we begin with ~\eqref{MOG2PartFuncChiralityFlip} where we replace $(\b,\m\b)=(2\p r_+,2\p r_-):$ 
\be
\begin{aligned}
Z(\b,\m;\t,\bar{\t})&=\sum_{m^\prime,n,w_1\in\mathbb{Z}}\sum_{q',\bar{q}',N,\bar{N}}\Biggl\{\Biggr\}
\times e^{2\p i\t_1\left[w_1(q-\bar{q})+nm^\prime+N-\bar{N}\right]}\times e^{-2\p\t_2\left[\frac{2s^2}{k-2}+N+\bar{N}\right]}\\
&\times e^{-2\p\t_2\left[-\frac{1}{4}-w_1\left(i(q-\bar{q})\frac{r_-}{r_+}+\frac{ im'}{r_+}+k nr_-\right)+\frac{k}{2}w_1^2+\frac{k}{2}n^2 (r_-^2+r_+^2)+i\frac{r_-}{r_+}m^\prime n+\frac{i n(q-\bar{q})}{r_+}(r_-^2+r_+^2)\right]}\,,
\end{aligned}
\ee
where the factors with braces account for the degeneracy as in that formula. We will focus on the second line in the following since that piece involves the currents of the underlying WZW model.

In this section, we will study the conformal weights obtained from the above q-series and show that it agrees with the expected structure for the BTZ black hole \cite{NS} once we take into account Wick rotation. 

The spectrum of the Lorentzian BTZ black hole was first explored in \cite{NS}, where the conformal weights $L_0,\bar{L}_0$ where shown to be expressed in terms of the {\it hyperbolic} generators of \sl2. Thus, for the Euclidean BTZ as well, we expect 
\begin{subequations}\label{L0pmL0barEBTZ}
\begin{align}
    L_0+\bar{L}_0&=\frac{1}{k}\biggl[-(J^{(2)}_{+,0})^2+(J'^{(2)}_{+,0})^2-(J^{(2)}_{-,0})^2+(J'^{(2)}_{-,0})^2\biggl]+\frac{2s^2+\frac{1}{2}}{k-2}+N+\bar{N}\,,\label{L0plusL0barEBTZ}\\
    L_0-\bar{L}_0&=\frac{1}{k}\biggl[-(J^{(2)}_{+,0})^2+(J'^{(2)}_{+,0})^2+(J^{(2)}_{-,0})^2-(J'^{(2)}_{-,0})^2\biggl]+N-\bar{N}\label{L0minusL0barEBTZ}\,.
\end{align}
\end{subequations}
with different sign choices compared to Thermal AdS \eqref{L0pmL0barTAdS3}. The Euclidean nature will again mean that the eigenvalues $J^{(2)}_{\pm,0}$ and $J'^{(2)}_{\pm,0}$ will not be real.

A choice of current eigenvalues satisfying these relations with $L_0\pm\bar{L}_0$ read off from~\eqref{MOG2PartFuncChiralityFlip} is
\begin{subequations}\label{CurrentsFromPartFuncEBTZ}
\begin{align}
&\begin{aligned}
    J^{(2)}_{\pm,0}&=\left[\frac{\pm m^\prime+ i(q-\bar{q})\D_\pm}{2r_+}\right]\pm\frac{k}{2}(-ir_-n+iw_1)\\
    &=\frac{-1}{2\D_\mp}\left[\frac{\mp m^\prime\D_\mp-i(q-\bar{q})(r_+^2+r_-^2)}{r_+}\right]\pm\frac{k}{2}(-ir_-n+iw_1)\,,\\
\end{aligned}\\
&\begin{aligned}
    J'^{(2)}_{\pm,0}&=J^{(2)}_{\pm,0}-\frac{k}{2}\left(\pm iw_1-n\D_\mp\right)\\
    &=\frac{-1}{2\D_\mp}\left[\frac{\mp m^\prime\D_\mp-i(q-\bar{q})(r_+^2+r_-^2)}{r_+}\right]+\frac{k}{2}r_+n\,,
\end{aligned}
\end{align}
\end{subequations}
where $\D_\pm\coloneqq r_+\mp ir_-$. At this stage, there are four integers: $m'$, $(q-\bar{q})$, $n$ and $w_1$. We expect these to be related to the winding and momentum quantum numbers along the $t_{\text{BTZ}}$ and $\f_{\text{BTZ}}$ directions of the Black hole geometry.

To interpret these currents, we use coordinates $(\hat{t},\hat{\f})$ which are rotated with respect to the TAdS coordinates $(\textcolor{red}{-i}t_{\text{global}},\f)=(-\hat{\f},\hat{t})$. Euclidean BTZ coordinates are {\it defined} by the requirement that the identifications inherited from TAdS map to the EBTZ identifications: 
\begin{subequations}\label{IndentificationsEBTZ}
\begin{align}
    \hat{t}&\sim\hat{t}+2\p &\Longleftrightarrow&&(t_{\text{EBTZ}},\f_{\text{BTZ}})&\sim(t_{\text{EBTZ}},\f_{\text{BTZ}})+\frac{2\p(r_+,r_-)}{r_+^2+r_-^2}\,,\\
    (\hat{t},\hat{\f})&\sim(\hat{t},\hat{\f})-2\p(r_-,r_+)&\Longleftrightarrow&&(t_{\text{EBTZ}},\f_{\text{BTZ}})&\sim(t_{\text{EBTZ}},\f_{\text{BTZ}})-(0,2\p)\,.
\end{align}
\end{subequations}
The EBTZ coordinates $(t_{\text{EBTZ}},\f_{\text{BTZ}})$ are then obtained as 
\begin{equation}\label{EBTZtoTADS}
    \hat{t}\coloneqq r_+t_{\text{EBTZ}}+r_-\f_{\text{BTZ}}\,,\qquad\hat{\f}\coloneqq r_+\f_{\text{BTZ}}-r_-t_{\text{EBTZ}}\,.
\end{equation}

The eigenvalues~\eqref{CurrentsFromPartFuncEBTZ} are evaluated in the spectrally flowed sector with $(\hat{t},\hat{\f})\xmapsto{\s\mapsto\s+2\p}(\hat{t},\hat{\f})+w_1(2\p,0)-n(2\p r_-,2\p r_+)$. This follows from the relation between $J'^{(2)}_{\pm,0}$ and $J^{(2)}_{\pm,0}$. In this sector, we may evaluate the target space momenta using the action~\eqref{WZWAction} with $(\textcolor{red}{-i}t_{\text{global}},\f)=(-\hat{\f},\hat{t})$ (which equals the EBTZ sigma model action) as before:
\begin{subequations}
\begin{align}
    -\Pi_{\hat{\f}}&\coloneqq\int_0^{2\p}\mkern-18mud\s\;\frac{\del\mathcal{L}}{\del(\del_\t\hat{\f})}=-J'^{(2)}_{+,0}+J'^{(2)}_{-,0}-kC(iw_1\red{-i}r_-n)\\
    &=-\frac{m'}{r_+}-\frac{\red{i}r_-}{r_+}i(q-\bar{q})-kCiw_1\red{+i}kCr_-n\,,\\
    \textcolor{red}{-i}\Pi_{\hat{t}}&\coloneqq i\int_0^{2\p}\mkern-18mud\s\;\frac{\del\mathcal{L}}{\del(\del_\t \hat{t})}=-J'^{(2)}_{+,0}-J'^{(2)}_{-,0}-k(C-1)r_+n=-i(q-\bar{q})-kCr_+n\,.
\end{align}\label{MomentaTF}
\end{subequations}
We may now verify that target space momenta are correctly quantised for $C=0$ i.e.,
\begin{equation}
    \Pi_{\hat{t}}=q-\bar{q}\in \mathbb{Z}\,,\qquad\Pi_{\f_{\text{BTZ}}}=-r_-\Pi_{\hat{t}}+r_+\Pi_{\hat{\f}}=m'\in\mathbb{Z}\,.
\end{equation}
with $m'$ being the angular momentum as appropriate for a point-particle.

The choice $C=0$ for the EBTZ action matches what would be obtained from the Thermal $AdS_3$ action by $(\textcolor{red}{-i}t_{\text{global}},\f)=(-\hat{\f},\hat{t})$. It implies that $B_{\hat{t}\hat{\f}}\propto r^2-r_+^2$ and hence vanishes at $r^2=r_+^2$ as expected.

The winding number along the angular direction $\f_{\text{BTZ}}$ is $n$ and that along the Euclidean time direction $\hat{t}$ is $w_1$. As in ${\text{TAdS}}_3$, this is clear from the $\s\mapsto\s+2\p$ transformation mentioned below~\eqref{EBTZtoTADS}. This means that $n$ labels the twisted sectors of the orbifolding that produces the Euclidean BTZ from $\mathbb{H}^3$ CFT since the orbifolding identification leads to the periodicity of the $\f_{\text{BTZ}}$-coordinate.

One key difference between the EBTZ and thermal $AdS_3$ interpretations of the same partition function~\eqref{MOG2PartFuncChiralityFlip} is that the EBTZ interpretation~\eqref{CurrentsFromPartFuncEBTZ} involves the hyperbolic generators $J^{(2)}_{\pm,0}$ rather than the elliptic ones. Because of this,~\eqref{L0pmL0barEBTZ} implies
\be
L_0=\frac{-(j-\frac{1}{2})^2+\frac{1}{4}}{k-2}-i\O_+(j+q)+\frac{k}{4}\O_+^2+N\label{eqn:RepNoIdentify}
\ee
(and similarly $\bar{L}_0$) using the facts that
\be
J^{(2)}_{\pm,0}=\pm i(j+q_\pm)\,,\qquad J'^{(2)}_{\pm,0}=\pm i(j+q_\pm)-\frac{k}{2}\O_\pm\,,
\ee
where $(q_+,q_-)\coloneqq(q,\bar{q})$, $\O_\pm\coloneqq\pm iw_1-n\D_\mp$ and $j\coloneqq j^{n,m'}_{q,\bar{q},w_1}$ with $(\b,\m\b)=(2\p r_+,2\p r_-)$ i.e,
\be
j\coloneqq j^{n,m'}_{q,\bar{q},w_1}=\frac{k}{2}w_1-\frac{1}{2}\left(i(q-\bar{q})\frac{r_-}{r_+}+\frac{im'}{r_+}-knr_-\right)-\frac{1}{2}(q+\bar{q})\,.
\ee
Here,~\eqref{eqn:RepNoIdentify} is different from~\eqref{eqn:RepIdentify} in the sign of the last term $+\frac{k}{4}\O_\pm^2$ in $L_0$ and $\bar{L}_0$, which is a consequence of the hyperbolic generators being diagonalised. This prevents us form reinterpreting a shift of the spatial winding number $n$ as a change of $j$ and $N$. Hence, differently spectral flowed discrete series representations are not to be identified in EBTZ. This agrees with our previous analysis~\cite{Nippanikar} for Lorentzian BTZ, where this representation theoretic fact could also be seen at the classical level through discrete symmetries of BTZ geodesics.

\subsection{Comparison of Partition functions}
\label{COMP}
In this section, we will compare the torus partition functions of Thermal AdS and that of Euclidean BTZ at the same temperature, chemical potential and boundary volume. We will do this by first writing the `torus factor' in the partition function appearing in the second line of \eqref{ThePartFun} in terms of the the boundary modular parameter $T$ and the Kahler parameter $\r_2.$ For the comparison to be meaningful, we will ensure that the boundary 
metrics are equal.
To this end, we must use the first form for the boundary metric of the EBTZ. As we have shown, this means that the boundary torus has a rescaled volume eqn 1.20.  We will see below that this rescaling can be obtained from the partition function as well. We begin by reminding ourselves that the partition function of \cite{MO2} has been computed assuming $\r_A=l.$

 We will rewrite the partition function eqn.(\ref{ThePartFun}) of TADS by expressing the chemical 
potential and temperature in terms of the Modular parameter $T_A$
and a complex Kahler parameter defined by $\r=\r_1+i\r_2$. Here,  $\r_1=J_A$ is determined by the non-normalizable mode of the B-field and $\r_2$ is the volume parameter of the boundary torus as determined from the non-normalizable modes of the metric. From the earlier results \eqref{B-fielddata}, we see that $\frac{\r_1}{\r_2}=i$ for both TADS and  EBTZ.

Thus,  substituting $\frac{\b\m}{2\pi}=\tilde\m\tilde\b=T_1$ and $ \frac{\b}{2\pi}=\tilde\b=T_2$ in eqn.(\ref{ThePartFun}), we obtain: 
\begin{multline}
\S_{TADS}(T)= \sum_{m',n,w_1,a} \t_2 \exp\biggl\{ -\pi \t_2 \bigg[-\frac{2\rho_1}{\rho_2} \left(\frac{T_1}{T_2}\left((s_1+w_1)a - m'n \right) +  \frac{m'(s_1+w_1)}{T_2}-\frac{|T|^2}{T_2}na \right)  \\ +
\frac{\rho_2}{T_2} \left|(s_1+w_1)-nT \right|^2\bigg]+
2\p i\left[\t_1(nm^\prime+a(s_1+w_1))- as_2)\right] \biggl\} \label{SigmaTAds}
\end{multline}
Similarly, we can obtain the EBTZ partition function in terms of the modular parameter. To this end, we first replace the orbifold parameters $(\b,\m\b)\to(r_+,r_-)$ in the TADS partition function~\eqref{ThePartFun} which gives us the EBTZ partition function. We then re-express $r_\pm$ using the modular parameter ${T}_B=-1/(-r_-+ir_+)$ and $\r$ as determined earlier \eqref{TempModBTZ} to get:
\begin{multline}
     \S_{EBTZ}(T)=\exp\biggl\{ -\pi \t_2 \bigg[ \frac{2\rho_1}{\rho_2} \left(\frac{\tilde{T}_1}{\tilde{T}_2}\left(\blue{-}(s_1+w_1)a + m'n \right) \blue{-}  \frac{m'(s_1+w_1)|\tilde{T}|^2}{\tilde{T}_2}+\frac{na}{\tilde{T}_2} \right)  \\ + 
 \frac{\rho_2}{\tilde{T}_2}|(s_1+w_1)\blue{-}\frac{n}{\tilde{T}}|^2 \bigg]\bigg]+
 2\p i\left[\t_1(nm^\prime+a(s_1+w_1))- as_2)\right] \biggl\} .\label{SigmaEBTZ}
\end{multline}

We can compare the TAdS and EBTZ {\it partition functions}. As shown in \cite{Malstrom}, EBTZ partition function $\S_{EBTZ}(T)$ with modular parameter $T$ must be compared with $\S_{TADS}(-\frac{1}{T})$. 
However, if we make the replacement, $T \to -\frac1T$ in the TADS partition function \ref{SigmaTAds}, the expression we get is not the same as the one above \ref{SigmaEBTZ}:
\begin{multline}
\S_{TADS}(-\frac1T)= \exp\biggl\{ -\pi \t_2 \bigg[\red{-}\frac{2\rho_1}{\rho_2} \left(\frac{T_1}{T_2}\left(-(s_1+w_1)a + m'n \right) +  \frac{m(s_1+w_1)|T|^2}{T_2}-\frac{na}{T_2} \right)  \\ +
\frac{\rho_2\red{|T|^2}}{T_2} \left|(s_1+w_1)\red{+}\frac{n }{T} \right|^2\bigg]+
2\p i\left[\t_1(nm^\prime+a(s_1+w_1))- as_2)\right] \biggl\} 
\end{multline}
where the differences are highlighted in red. On the other hand, as shown previously, the volume parameters of TADS and EBTZ are related as $\r\to\frac{\r}{|T|^2}$. In addition, recall that the TADS partition function of \eqref{ThePartFun} was for a chemical potential $-\m$ (section \ref{sec:EBTZViewpoint}). This requires us to flip the sign of $T_1$ in the BTZ expression before comparing. Making this change, and flipping the signs of $n,m'$, we get the EBTZ partition function. 

This agrees with the observations made earlier that the EBTZ geometry requires a $\pi/2$ rotation, since that amounts to a sign flip of the $J_-$ current. This does not affect $L_0, \bar{L}_0.$ 
Thus, the BTZ black hole is thus not merely the S-dual of the AdS but it also requires a rotation and an additional rescaling of the volume $\rho \to \frac{\rho}{|T|^2}$. This agrees with our analysis in the first section.

\section{The Lorentzian spacetimes}
The spectrum of states for strings on the Lorentzian spacetimes of $AdS_3$ and BTZ black holes have been presented in \cite{MO1} and \cite{Nippanikar}. In this section, we will show that the CFT spectra can be obtained from the {\it Euclidean} partition functions of the previous section by Wick rotating and decompactifying the time circle. It is not surprising that we are able to decompactify -- after all $AdS_3$ itself requires us to decompactify the \sl2 WZW model. Further motivation to expect these Wick rotations arises from our observations in Sections~\ref{sec:TAdSViewpoint} and~\ref{sec:EBTZViewpoint}. In particular, the differently spectral flowed discrete series representations are identified in thermal $AdS_3$ and not identified in EBTZ, in agreement with their Lorentzian versions. The surprise is that the Wick rotation proceeds so smoothly and gives us the Lorentzian spectrum quite painlessly. 


\subsection{Wick Rotation of the TADS partition function}

In this section, we will see that we can directly Wick rotate the Thermal AdS partition function to the Lorentzian AdS partition function. The partition function for the universal cover of \sl2 as written by Israel et.al. \cite{Israel} is:
\begin{eqnarray}
Z_{SL(2,\mathbb{R})} \left(\t , \bar{\t} \right) &=& 4 \sqrt{\tau_2} (k-2)^{3/2} \int_0^1 d^2 s \,
\int_0^1 d^2 t \frac{ \mathrm{e}^{\frac{2\pi}{\tau_2} (\mathrm{Im} (s_1
\tau -s_2))^2 } }{\left|\vartheta_1 (s_1 \tau -s_2 |\tau )
\right|^2}\times  \nonumber \\
& &\times  \sum_{m,w,m',w'\in \mathbb{Z}} \zeta\oao{w + s_1 - t_1
}{m +s_2 - t_2 }\left(k \right)  \zeta\oao{w' + t_1 }{m' + t_2
}\left(-k \right). \label{eqn:miunisl2}
\end{eqnarray}
where 
\be 
\zeta\oao{\omega}{\mu}\left(k\right) = \sqrt{\frac{k}{\tau_2}}
\exp\left[ -\frac{\pi k}{\tau_2}\left|\omega \tau - \mu\right|^2\right]
\label{eqn:zetadefn.}
\ee 
The above partition can be expanded (following \cite{Israel}) into the same form as that in Eqn\eqref{MOG2PartFuncChiralityFlip} so that we can compare the spectrum. This is
\be 
\begin{aligned}   
Z &= \frac{4}{\pi} \left(k-2 \right) \int_{-\infty}^{\infty} ds \int_0^1 d^2 t \ \sum_{w,w'} \sum_{n,n'} \sum_{\substack{q,\bar{q} \\ N, \bar{N}}} \d_{q-\bar{q}, n} 
\mathrm{exp} \left[2 \pi i \t_1 \left( n \left(w - t_1 \right) + n' \left(w' + t_1 \right) + N - \bar{N} \right)  \right]  \\
&\mathrm{exp}\left[2 \pi t_2 (n - n')  \right] \left[\frac{1}{2is + q + \bar{q} + 1 + k\left(w-t_1 \right) } - \frac{1}{2is + q + \bar{q} - 1 + k\left(w-t_1 \right) }  \right] \\ 
&\mathrm{exp} \left[-2 \pi \t_2 \left( \frac{2s^2 + 1/2}{k-2} + \frac{n^2}{2k} - \frac{n'^2}{2k} - \frac{k}{2}\left(w'+t_1 \right)^2 + \frac{k}{2} \left(w-t_1 \right)^2 + N + \bar{N} - \frac{3k}{12 \left(k-2 \right) } \right) \right] 
\end{aligned}
\ee 
We will compare this Lorentzian spectrum with the Euclidean Thermal AdS spectrum \eqref{MOG2PartFuncChiralityFlip}. To begin, we list the coefficients multiplying $\t_1$ in the exponent
\bea
TADS&:& \left[w_1(q-\bar{q})+nm^\prime+N-\bar{N}\right]\\
LADS&:& \left([ n \left(w - t_1 \right) + n' \left(w' + t_1 \right) + N - \bar{N} \right]
\eea
and $\t_2$
\begin{small}
\bea
TADS&:& \left[\frac{2s^2}{k-2}-\frac{1}{4}-w_1\left(i(q-\bar{q})\m+\frac{2\p i}{\b}m^\prime+\frac{\m k\b n}{2\p}\right)+\frac{k}{2}w_1^2+\frac{k}{2}\left(\frac{\b n}{2\p}\right)^2(\m^2+1)\right]\\
LADS&:& 
\left[ \frac{2s^2 }{k-2}-\frac{1}{4} + \frac{n^2}{2k} - \frac{n'^2}{2k} - \frac{k}{2}\left(w'+t_1 \right)^2 + \frac{k}{2} \left(w-t_1 \right)^2 \right]
\eea

\end{small}
In the Lorentzian ADS, we need to set $n=n'$ because of the constraint arising from $t_2$ integration.

Starting with the TADS expression, we will take the zero temperature $\b\to\infty.$
From the last term in TADS we see that $n$ must be set to zero.  Also we replace $\frac{2\p i}{\b}m^\prime=E$ a real number which involves a Wick rotation. 
Then, omitting obviously common terms in the $\t_2$ expression, we compare 
\bea
TADS&:& \left[-w_1\left(E+\frac{k}{2}w_1^2\right)\right]\\
LADS&:& 
\left[- \frac{k}{2}\left(w'+t_1 \right)^2 + \frac{k}{2} \left(w-t_1 \right)^2 \right]
\eea
Let us now consider the $\t_1$ factors which requires us to compare
\be
w_1(q-\bar{q})\  {\rm and }\  n(w+w')
\ee
Thus, we can set $w_1 = -(w+w')$ since we know that $(\bar{q}-q)$  is the angular momentum $n$. Finally, if $E=-k(w'+t_1)$, the exponents in the partition function of Thermal AdS and Lorentzian AdS will be identical. Thus, starting with TADS, we determine $w'$ from $E$ as the integral part. Then we write $-w_1=w'+w$ - and we get LADS. 

Using these identifications, we can compare the density of states of the two systems. The expression for the factor that gives rise to the density of states in TADS rewritten using the above identifications is 
\be
\Biggl[\int_{-\infty}^\infty\biggl(\tfrac{ds/\p}
{2is-kw_1+E+1+q+\bar{q}}-\tfrac{ds/\p}{2is-kw_1+E-1+q+\bar{q}}\biggr)+\sum_{s\in D_{q,\bar{q},w_1}^{n,m^\prime}}\Biggr]
\ee
which we compare with
\be
\Biggl[\frac{1}{2is + q + \bar{q} + 1 + k\left(w-t_1 \right) } - \frac{1}{2is + q + \bar{q} - 1 + k\left(w-t_1 \right) }  \Biggr]
\ee
which are equal once we use the identfications. Because the denominators become the same, the discrete series poles are at the correct location. Note that because of the real number $t_1$ in the denominator, the j-values of the discrete series are not integral (as is correct for the universal cover). 

Thus, in conclusion we have shown that the spectrum of Lorentzian $AdS_3$ could have been extracted from the partition function of Thermal $AdS_3$ as a zero temperature limit and Wick rotating.

\subsection{The other Wick rotation}

In this section, we will try to show that the spectrum of the Lorentzian BTZ black hole as worked out in \cite{Nippanikar} can be obtained from the Euclidean BTZ black hole by Wick rotation. 
In this section, we will explicitly use suffixes $L$ for Lorentzian BTZ quantities and $E$ for those of the Euclidean BTZ.

Starting with the EBTZ currents~\eqref{CurrentsFromPartFuncEBTZ}, we first Wick rotate $r_-\to ir_-$ and set $w_1=0$, which is the winding number along the Euclidean time direction $\hat{t}$, to get
\begin{subequations}
\begin{align}
    J^{(2)}_{\pm,0}
    &=\frac{-1}{2\D_\mp}\left[\frac{m'r_-+kr_+r_-^2n_E-i(q-\bar{q})_E(r_+^2-r_-^2)}{r_+}\pm\left(-m'-kr_+r_-n_E\right)\right]\,,\\
    &=\frac{-1}{2r_+\D_\mp}\left[
    (m'+kr_+r_-n_E)(r_-\mp r_+)-i(q-\bar{q})_E(r_+^2-r_-^2)\right]
    \\
    J'^{(2)}_{\pm,0}
    &=J^{(2)}_{\pm,0}+\frac{k}{2}n_E\D_\mp\,,
\end{align}
\end{subequations}
where $\D_\pm\coloneqq r_+\pm r_-$ and the subscript $E$ is placed to help differentiate the EBTZ quantum numbers from those of Lorentzian BTZ in what follows.

These may be compared with the form of current eigenvalues for Lorentzian BTZ presented in \cite{Nippanikar}, which is
\begin{subequations}
\begin{align}
    \tilde{J}^{(2)}_{\pm,0}
    &=\frac{-k}{2\D_\mp}\left(\tilde{E}\pm \tilde{L}\right)\,,\\
    J'^{(2)}_{\pm,0}&=\tilde{J}^{(2)}_{\pm,0}-\frac{k}{2}\O\D_\mp
\end{align}
\end{subequations}
where $k\tilde{E}$ and $k\tilde{L}$ are related to the spacetime energy $kE\in\mathbb{R}$ and angular momentum $kL\in\mathbb{Z}$ of the string via
\begin{align}
    E&=\tilde{E}+(r_+^2+r_-^2)\O\,,&L&=\tilde{L}-\O r_+r_-\,.
\end{align}
The comparison makes it clear that the Wick rotation maps the Euclidean quantum numbers to the Lorentzian ones via
\begin{align}
    m'&=-kL\,,&n_E&=-\O\,,&
    -i(q-\bar{q})_E&=\frac{k(\tilde{E}r_++r_-\tilde{L})}{r_+^2-r_-^2}=-\tilde{J}^{(2)}_+-\tilde{J}^{(2)}_-
\end{align}
It satisfactory that the Euclidean momentum $\Pi_{\hat{t}}=(q-\bar{q})_E$ Wick rotates to a (purely imaginary) {\it continuous} variable, while the angular momentum $\Pi_{\f_{\text{BTZ}}}=m'$ and the spectral flow $n_E$ remain discrete after Wick rotation.   

The Wick rotation of the density of states on the other hand does not proceed in an as intuitively reasonable manner:
\bea
(q+\bar{q})_E+ik[\frac{(\tilde {E}r_-+\tilde{L}r_+)}{r_+ ^2-r_-^2}]&=(q+\bar{q})_L-k(\tilde E+\frac{1}{2}(M+1)\O)\\
(q+\bar{q})_E+i(\tilde{J}^{(2)}_+ -\tilde{J}^{(2)}_-)&=(q+\bar{q})_L-\frac{k}{2}(M+1)\O+\tilde{J}^{(2)}_+\D_- +\tilde{J}^{(2)}_-\D_+
\eea




\section{The B-field}

If we have a gauge field $A_a$ in the bulk, its boundary value acts as a source to a boundary operator $J^\m$ in the gauge $A_r=0.$ 
The non-normalizable modes of $A_\m$, thus appear in the form  $\int d^{d}x A_\m J^\m$ in the boundary theory Hamiltonian.  Under the residual gauge symmetry, the boundary term changes as
$\int \del_\m \L J^\m=-\int \L \del_\m J^\m.$  Since gauge invariance ought to be a symmetry, requiring the above variation to vanish obtains {\it local} current conservation. Note that this involves a gauge transformation parameter $\L(x)$ which does not vanish at infinity $\L(x)=\L(r=\infty, x).$ 

If we have a B-field in the bulk, then it can couple to two kinds of boundary operators - to vector operators in the manner of $B_{\r\m} J^\m$  and to a  rank two tensor operator in the CFT as $B_{\m\n}J^{\m\n}.$ 

The first kind of terms gauge transform as $(\del_\r \L_\m-\del_\m\L_r)J^\m$. If $\del_\r\L_\m$ survives at the boundary, this means that $\L_\m(\r)$ does not vanish at the boundary, i.e., it is not a small gauge transformation, and hence is not an allowed gauge transformation in the bulk. Thus, the residual gauge symmetry of the B-field is $\L_r(x)$ which gives rise to current conservation as in the case of the bulk vector field.  

Under a gauge transformation of $B$ the total gauge variation of the {\it tensor} coupling evaluates to 
\be
\int d^dx \L^\n \del^\m(J_{\m\n})=0
\ee
where $\L$ is the gauge parameter. Requiring that the above variation vanishes gives an identity satisfied by the dual operator 
\be
\del_\m J^{\m\n}=0
\ee
Since $J^{\m\n}$ is antisymmetric, this equation cannot be interpreted as a continuity equation and therefore it does not give a regular conservation law. However, this is a generalized symmetry as presented in say, \cite{McGreevy}. 

We write out this integral for the boundary of {\it Lorentzian} $AdS_3$ to get 
\be\int d\f (\L_\f J^{t\f})|_{t=a} ^{t=b}+ \int dt (\L_t J^{\f t})|_{\f=0} ^{\f=2\pi}\ee
This quantity must vanish for all choices of $\L_{t,\f}.$ 
Requiring this quantity to vanish in Lorentzian BTZ geometry forces $J_{t\f}=0$ - simply take $\L_\f=\delta(\f-\f_0).$ 

In the Euclidean BTZ case, the above integral is over a torus on which the vector field $\L_\n$ is not allowed to be topologically non-trivial since it is a gauge tranformation parameter. This certainly seems to allow $J_{\m\n}=J_{BH}\neq 0$ thereby explaining the non-normalizable mode computed from the bulk.

We also note that in the Lorentzian case, the B-field vanishes at $r^2=M$ \cite{Esko}. 
From the Lorentzian BTZ point of view, 
the finite gauge transformation that is required to take us from one gauge  $C=0$ to another $C\neq0$ will translate into either a time dependent source $A_\f(t)$ or a non-trivial (i.e, non-periodic) boundary Wilson line $A_t(\f).$ Thus, we seem to be driven to the conclusion that Euclidean and Lorentzian BTZ are unrelated from the Boundary theory viewpoint.

\section {Boundary Modular Invariance}
In this section, we will discuss the {\it boundary} modular invariance of BTZ and Thermal AdS partition functions; that is, the modular invariance of \eqref{MOG2PartFuncChiralityFlip} with respect to the parameter $T$.  Since the dual theory is a CFT which lives on a torus defined by $T$ and $\r$, we can expect, based on the general properties of 2D CFTs, that physical quantities of the boundary theory are invariant under modular transformations of $T.$ While the worldsheet torus partition function of the bulk is not a physical object in string theory, it is nevertheless interesting to ask if we have modular invariance with respect to the spacetime parameters $T.$ 

We recall that a spacetime torus {\it compactification} of string theory is characterized by two modular parameters representing coordinates on the $O(d,d,R)$ moduli space \cite{Polchinski}. The first parameter is the complex structure of the torus $T=T_1+iT_2.$ The second parameter is a Kahler modulus which involves the volume of the torus and a NS-NS B-field $\r=B+i\r_2.$ The T-duality group $O(2,2,Z)$ then appears as separate modular invariances for $\r$  and $T.$ However, the TADS and EBTZ geometries are warped compactifications of the torus in which the torus volume varies along the AdS radial coordinate.  
Thus, we expect only a $SL(2,Z)$ Buscher \cite{Buscher} duality which changes the complex structure of the torus. This  T-duality of the bulk theory is valid order by order in the string coupling $g_s$ and hence, we expect that the one loop amplitude which is the integrated torus partition function (after including ghosts and internal CFTs) to be invariant under full modular group action on $T$. On the other hand, the bulk is a warped compactification on a torus and hence we do not expect modular invariance under $\r.$

We start with our observation that the TAdS and the EBTZ partition function~\eqref{ThePartFun} can be expressed in the form
\be
Z_g(\rho,T;\t) = \t_2  \sqrt{(k-2)}\int_0^{1} \mkern-18muds_1 \int_0^{1} \mkern-18muds_2 \frac{e^{\frac{2\pi}{\t_2} Im[s_1 \t - s_2]^2}}{\sqrt{\t_2}|\vartheta_1(s_1 \t - s_2|\t)|^2}\   \Sigma_g[s_2,s_1,T,\rho],
\ee
where $\S_g$ contains all the parameters referring to the boundary torus and the subscript $g$ refers to the geometries - $TAdS_3$ or $EBTZ$.

For $TAdS$, we have: 
\begin{multline}
\S_{TADS}(T)= \sum_{m',n,w_1,a} \t_2 \exp\biggl\{ -\pi \t_2 \bigg[-\frac{2\rho_1}{\rho_2} \left(\frac{T_1}{T_2}\left((s_1+w_1)a - m'n \right) +  \frac{m'(s_1+w_1)}{T_2}-\frac{|T|^2}{T_2}na \right)  \\ +
\frac{\rho_2}{T_2} \left|(s_1+w_1)-nT \right|^2\bigg]+
2\p i\left[\t_1(nm^\prime+a(s_1+w_1))- as_2)\right] \biggl\} \label{SigmaTAds2}
\end{multline}
Before we study modular properties with respect to $T$, it helps to perform Poisson resummation on $w_1,n$ in that order, with $y,u$ being the respective dual variables. This gives us the following expression
\bea
\S_{TADS}&=&\frac{1}{T_2k}e^{[2\pi i (as_2+s_1 y) -\frac{\pi}{k\t_2 T_2 ^2}(\bar{Y} T+\bar{U})(Y\bar{T}+U)]}\\
&=&\frac{\r^2_A}{4\r_2}e^{2\pi i (as_2+s_1 y)}\,\, e^{-\frac{\r^2_A\pi}{4\r_2\t_2T_2}( (\bar{Y} T+\bar{U})(Y\bar{T}+U))}\label{TorusForm}
\eea
where $Y=y+a\t$ and $U=u+m'\t$ and $T=T_1+i T_2$ and $\t=\t_1+\frac{\r_1}{\r_2}\t_2$ and $\bar{\t}=\t_1-\frac{\r_1}{\r_2}\t_2.$

In this manner of writing
\be
Z_{TADS}(T)=\left\{\frac{\sqrt{k-2}\ e^{\frac{2\pi}{\t_2} Im[s_1 \t - s_2]^2}}{\sqrt{\t_2}\ |\vartheta_1(s_1 \t - s_2|\t)|^2}\right\}
\,e^{2\pi i (as_2+s_1 y) }\, \left\{
\frac{(2\pi)^2}{\rho_2}e^{-\frac{\pi (2\pi)^2}{\rho_2 T_2 \t_2}(Y T+U)(\bar{Y}\bar{T}+\bar{U})}\right\}
\ee
where each factor in braces is modular invariant on the worldsheet by itself.
The second factor is the partition function of a spacetime torus with modular parameter $T$ and Kahler parameter $\r.$ It is coupled 
to the ``radial coordinate " which is represented by the first factor via the Lagrange multipliers $s_{1,2}.$ 
In particular, the modes $a,y$ that couple to $s_1,s_2$ in the Thermal AdS case \ref{sec:TAdSViewpoint} are related to the momentum and winding along the $\f$ circle which contracts to zero size at $r=0.$

\subsection{Poincar\'e series}
Under $T\to T+1$, the above expression is easily seen to be invariant if we also replace $w_1\to w_1+n, m'\to m'-a.$ 
Under a S-modular transformation $T\to -\frac{1}{T}$, we get 
\be
Z_{TADS}(T)=\sqrt{k-2}\frac{e^{\frac{2\pi}{\t_2} Im[s_1 \t - s_2]^2}}{\sqrt{\t_2}|\vartheta_1(s_1 \t - s_2|\t)|^2}\,e^{2\pi i (ms_2+s_1 u) }\, 
\frac{(2\pi)^2}{\rho_2}e^{-\frac{\pi (2\pi)^2}{\rho_2 T_2 \t_2}(Y \bar{T}+U)(\bar{Y}T+\bar{U})}.
\ee
As discussed in \ref{COMP}, under this transformation - the partition function we obtain is interpreted as that of a Euclidean BTZ black hole. In particular, the time circle contracts to zero size. From this partition function, we can deduce, by expanding in a q-series, that $m',u$ are related to the energy and momenta along the time circle. This motivates us to interpret the modes which couple to $s_{1,2}$ as defining a contractible cycle in the bulk.

Under a general modular transformation $T\to\frac{AT+B}{CT+D}$, the torus factor in the partition function becomes
\bea
&=\frac{1}{\r_2}e^{2\pi i (as_2+s_1 y)}\,\, e^{-\frac{\pi}{\r_2\t_2T_2}( (Y(A\bar{T}+B)+U(C\bar{T}+D))(\bar{Y}(AT+B)+\bar{U}(CT+D)))}\\
&=\frac{1}{\r_2}e^{2\pi i (as_2+s_1 y)}\,\, e^{-\frac{\pi}{\r_2\t_2T_2}( ((AY+CU)\bar{T}+BY+UD)((A\bar{Y}+C\bar{U})T+(B\bar{Y}+\bar{U}D))}
\eea
We can define a new U and a new Y variable $Y'=AY+CU \qquad U'=BY+UD$ which written in matrix form
\be
\begin{pmatrix}a'\\y'\\m'\\u'\end{pmatrix}=\begin{pmatrix}A &0&C&0\\0&A&0&C \\B&0&D&0\\0&B&0&D\end{pmatrix}
\begin{pmatrix}a\\y\\m\\u\end{pmatrix}
\ee
allows us to see that the partition function changes to 
\be
e^{2\pi i\left[(a'D-Cm')s_2+s_1 (-Cu'+Dy')\right]}e^{-\frac{\pi}{\r_2\t_2T_2}( (Y' \bar{T}+U')(\bar{Y'}T+\bar{U'}))}
\ee
Note that note that only $C,D$ elements of the modular matrix appears in the partition function. To obtain a boundary modular invariant partition function, we can sum over the integers $A,B,C,D$ - but this naive sum will be infinite.

We therefore proceed as follows. 
It is manifest that we have modular invariance with respect to $T.$ Performing $T\to T+1$ can be cancelled by 
\be
U\to U-Y,\ \ \  \begin{pmatrix}A&B\\C&D\end{pmatrix}\to \begin{pmatrix}A&B+A\\C&D+C\end{pmatrix} \label{Tm}
\ee
while $T\to -\frac{1}{T}$
requires 
\be
U\to Y\ \  Y\to -U,\ \ \  \begin{pmatrix}A&B\\C&D\end{pmatrix}\to \begin{pmatrix}-B&A\\-D&C\end{pmatrix}\label{Sm}
\ee
Suppose $C,D$ are mutually prime. Then, Bezout's identity 
\be
1=C(x-kD)+D(y-kC)
\ee
states that given one solution $x,y$ of $1=Cx+Dy$ (which certainly exists) all solutions are of the above form. We can interpret $A=x-kD$ and $B=y-kC$ giving us an $SL(2,\mathbb{Z})$ matrix.  Since $A,B$ do not appear in the partition function, summing over $k$ will give us an infinity. Thus, we can simply set $k=0$ and sum over $C,D$ mutually prime to obtain an invariant expression.  From eqns \eqref{Tm} and \eqref{Sm}, we see that this set is closed under $T\to T+1$ and $T\to -\frac1T.$

The string theory torus amplitude is obtained by including an internal CFT and the worldsheets ghosts and finally integrating over the worldsheet torus moduli.

At this stage, we should review our expectations about boundary modular invariance. We expect the following two properties to hold for the boundary partition function
\begin{itemize}
\item It should be invariant under $T\to\frac{AT+B}{CT+D}$ 
\item
It should exhibit a factorization property 
\[Z(T)=\sum_n F_N(T)\bar{F}_N(\bar T).\]
\end{itemize}
These expectations originate from the conformal invariance of the dual. 

The string theory torus amplitude can be regarded as the one-loop contribution to the {\it free energy} of the boundary theory. 
This is because, by analogy with field theory, it is the sum over one loop, connected, vacuum diagrams. 

Thus, we may write a boundary modular invariant expression for the partition function by first exponentiating the integrated amplitude and then summing over $SL(2,\mathbb{Z})$ images:  
\be
\begin{aligned}
Z_{CFT}(T,\r)=e^{-\b F_{CFT}(T,\r)}=\sum^{\prime}_{C,D}\exp\biggl[\int\frac{d^2\t}{\t_2}\int_0 ^1 d^2s \frac{(2\pi)^2\sqrt{k-2}}{\rho_2\sqrt{\t_2}} \frac{e^{\frac{2\pi}{\t_2} Im[s_1 \t - s_2]^2}}{|\vartheta_1(s_1 \t - s_2|\t)|^2}\times\\
Z_{int}(\t)\, Z_{gh}(\t)
\sum_{a'm',y',u'}e^{2\pi i\left[(a'D-Cm')s_2+s_1 (-Cu'+Dy')\right]}e^{-\frac{\pi}{\r_2\t_2T_2}( (Y' \bar{T}+U')(\bar{Y'}T+\bar{U'}))} \biggl]\, 
\label{BoundPartFun}
\end{aligned}
\ee
where the prime on the summation restricts $C,D$ to mutually prime values. Each term in this summation is an AdS background or BTZ type background which depends on the torus mode that couples to the radial mode $s$.  This series is akin to the Farey-tail expansion of the elliptic genus discussed in \cite{DMMV} (see also \cite{MM}). 

While this expression has the desired invariance, it does not manifestly exhibit the factorization property. Indeed, it is not clear that we even have a factorization property - especially since the summation and integration involves delicate issues of regularization and order of limits.

The quantity obtained is the stringy one loop contribution to the CFT effective action (at finite temperature). We can divide the log of the partition function by the temperature $T_2$ to obtain the Free energy. 
We emphasize that this answer is obtained by keeping the prefactor $\r_2$ untouched. If we unpack $\r_2=kT_2$ and keep $k$ fixed, then it is not clear that we can achieve modular invariance.

\subsection{Phase transition}
Since the $Z_{CFT}$ is boundary modular invariant, we may interpret the partition function along the lines of \cite{DMMV}. 
Increasing the temperature, while keeping the chemical potential fixed, amounts to {\it rescaling} $T\to \l T.$ Beyond a certain value of $\l$, we will exit a fundamental domain for $T.$  We then have to perform a modular transformation to return to the fundamental domain. In general, this changes the contractible cycle. If the starting point was understood as a tower of excitations over a TADS geometry (say, defined by the A-cycle of the boundary torus being contractible),  we will transit to a black hole {\it interpretation} since the contractible cycle will be a linear combination of the A- and B-cycles. Thus, phase transitions, if any,  can be expected to appear at the boundary of the fundamental domain of the modular parameter $T$.

This condition agrees with that we obtained from regular thermodynamics \ref{HP}. The Hawking-Page transition occurred past $\frac{\b^2(1+\m^2)}{4\pi^2}=1$ which expressed in terms of the modular parameter is precisely $|T|^2=1.$ This is one of the boundaries of the usual `keyhole' fundamental domain for the modular parameter $T$ of a torus in the UHP. On further increasing the temperature, the further modular transformations will almost never bring us back to the situation where the spatial  circle contracts (except maybe some number theoretic cases). So, there are no further phase transitions. 

This view of the above partition function is complementary to the usual approach, wherein for each value of $T$, presumably a particular term in the summation over $C,D$  contributes dominantly. As we vary $T$, the same dominant term continues until we reach the edge of the fundamental domain. Then the dominant term changes and we get a phase transition.  In this view, we regard $T$ as a complex parameter (unrestricted by identifications). 


In either view, the black hole appears because we modified our interpretation. In the former approach the contractible changed its meaning while in the latter view the dominant term changed. This does not give us a physical (dynamical) picture of what produces the phase transition similar to the manner in which Landau-Ginzburg free energies carry no information about the physical interaction mechanisms driving a phase transition. Of course, since we are describing equilibria, a dynamical reason is not to be expected. 


\section{Summary}
In this article, we began by reworking the analysis in \cite{Malstrom} along the conventional lines of AdS/CFT correspondence. This approach highlights the role played by the volume parameter $\r_2$ of the boundary torus. We showed that the AdS orbifolding with parameter $\frac{-1}{T}$ produces a BTZ black hole with a rescaled boundary volume. This also allowed us to observe that the EBTZ and Thermal AdS differ in the sense that the VEV of the operator dual to the $B_{NS}$ field are not equal \cite{Berkooz:2007fe}.
Then, we revisited the CFT torus partition functions of Thermal AdS. We first showed how to manipulate the partition function so that all dependence on the temperature and chemical potential can be isolated. This form of the partition function then allowed us to show how the same partition function with reinterpreted parameters encodes the states of the Euclidean BTZ CFT. 
We also showed how Wick rotations can be performed on the Thermal AdS partition function which directly gave us the partition function of the Lorentzian AdS worked out earlier in \cite{Israel}. Somewhat surprisingly, the analogous Wick rotation of the BTZ works out painlessly and gives the spectrum argued for in \cite{Nippanikar}.
We were then able to write down a formula for the bulk partition function that is actually invariant under boundary modular transformations. 

The latter allowed us to conjecture that the Hawking-Page transition arises at the boundary of a fundamental domain for the boundary modular parameter. This 
transition seems to be quite insensitive to the internal CFT required in a critical string theory as well. It must be noted that the backgrounds appearing in the Poincar\'e series cannot be interpreted as
microstates since most terms are Euclidean black holes with finite entropy. 

We note that this Boundary modular invariance can be regarded in two distinct ways. From the perspective of pure 3D gravity,  which could have a boundary dual, the NS-NS B-field does not exist as a degree of freedom.
Thus, the modular invariance proof should not depend on the parameter $\r_{1,2}.$ In fact, there will be no {\it complex} boundary degree of freedom dual to $\r.$ 
On the other hand, if we regard the geometries as part of a string theory, then the normalizable mode of the B-field becomes a boundary datum and the boundary modular invariance will involve the Kahler modulus $\r=B_{NS}+i\r_2$ as well. 
These two points are nicely compatible with the Poincar\'e series form for the bulk partition function. 

From the vantage of AdS/CFT it has been argued that we could/should treat the cosmological constant as a boundary datum as well - dual to pressure in the boundary theory \cite{Karch}. We note that the the volume parameter $\r_2$ is the natural thermodynamic dual of pressure and is also proportional to the level $k$. We have seen that to ensure modular invariance, boundary S-transformations on the modular parameter $T$ must be supplemented by a rescaling of the cosmological constant as well. This is consistent with the \eqref{KahlerModuli} where we see that the volume parameters of EBTZ and TADS are rescaled by $|T|^2.$ 


As we have noted, the Lorentzian and the Euclidean BTZ differ not only by the Wick rotation but also by a large gauge transformation of the B-field. The bulk B-field gives rise to a generalized symmetry of the boundary system. In particular, the D1/D5 system which is a gauge theory living in 1+1 dimensions with fundamental matter must possess such symmetries and therefore interesting phases. The 1-form gauge symmetry of the B-field gives rise to a conservation law of a stringy operator in the boundary theory \cite{McGreevy}. It will be interesting to understand the difference between the states of the boundary theory corresponding the Lorentzian and Euclidean BTZ black holes from this viewpoint.

Another interesting calculation is to perform the integration over $\t$ in the boundary partition function \eqref{BoundPartFun} and thereby determine the dominant saddle, if any.
This will however not answer why we need to change the {\it interpretation} of the geometry (in terms of the contractible cycle) beyond the Hawking-Page transition. An idea in this context will be to consider the nature of the statistical fluctuations using a statistical analogy. In the Ising model, at high temperatures the low lying excitations are described by spin waves whereas the low temperature excitations are individual spins.  In terms of these effective degrees of freedom, we have statistically small fluctuations around the dominant saddle (`ground state') - which is a ``condensate'' of spins at low temperature and a condensate of ``spin-waves" at high temperature. Perhaps a similar expectation is reasonable - the effective excitations change across the Hawking-Page transition. At low temperatures, the momentum modes around the thermal circle are the important fluctuations while at high temperatures, the effective degrees of freedom involve some winding along a spatial direction as well. This picture will clearly not be relevant for the Lorentzian spacetimes.


Finally, yet another way of writing the torus factor \eqref{SigmaTAds2} is of some interest. String theory suggests that the last factor can be rewritten 
\be
e^{-\frac{\pi}{\r_2\t_2T_2}( (Y' \bar{T}+U')(\bar{Y'}T+\bar{U'}))}=e^{-V^T M V}
\ee
where the metric $M$ (on the `` charge lattice") transforms covariantly under the boundary modular transformations.
$M$ in our case, will be a $4\times 4$ dimensional matrix. We can imagine Wick rotating this matrix corresponding to Wick rotating the boundary time variable. It will be of interest to ask whether the Poincar\'e series survives the Wick rotation and explore its spacetime meaning.

\appendix

\section{Convention used in the thermal \texorpdfstring{$AdS_3$}{AdS3} partition function}\label{sec:AppendPathIntegral}
The aim of this appendix is to justify the choice $U_{n,m}=\frac{\b}{2\p}(\m-i)(n\t-m)$ in~\eqref{eqn:StartingPartFunc} which differs from that in~\cite{MO2} by complex conjugation. We begin with the action~\eqref{WZWAction} for the $SL(2,\mathbb{R})$ WZW model and perform the Wick rotation $-it_{\text{global}}=t\in\mathbb{R}$ to obtain the Euclidean $AdS_3$ action. Next, we switch from Lorentzian to Euclidean worldsheet using $(z,\bar{z})=(e^{ix^+},e^{ix^-})$. These changes result in $iS_{\text{WZW}}[g]=-S_E[g]$ where
\be
S_E[g]=\frac{k}{\pi}\int d^2z\;\bigl[\partial\r\bar{\partial}\r+\partial t\bar{\partial}t\cosh^2\r+\partial\f\bar{\partial}\f\sinh^2\r+i\sinh^2\r\left(\bar{\partial}t\partial\f-\bar{\partial}\f\partial t\right)\bigr]\label{eqn:EuclideanAction}
\ee
is the Euclideanised worldsheet action. We have set $C=0$ because of arguments presented in Sections~\ref{sec:TAdSViewpoint} and~\ref{sec:EBTZViewpoint}. Also, $d^2z\coloneqq\frac{dzd\bar{z}}{(-2i)}$ so that $\int d^2z=4\pi^2\t_2$ is the area of the worldsheet torus defined by $(z,\bar{z})\sim(z+2\pi,\bar{z}+2\pi)$ and $(z,\bar{z})\sim(z+2\pi\t,\bar{z}+2\pi\bar{\t})$. The integrand in~\eqref{eqn:EuclideanAction} may be rewritten in terms of
\be
v=e^{i\f}\sinh\r\,,\qquad\bar{v}=e^{-i\f}\sinh\r\,,\qquad\xi=t-\ln\cosh\r\,,
\ee
upto a divergence of the vector field $(X^z,X^{\bar{z}})=(-i\bar{\partial}\f\ln\cosh\r,i\partial\f\ln\cosh\r)$ as
\be
\partial\xi\bar{\partial}\xi+\left(\bar{\partial}\bar{v}+\bar{\partial}\xi\bar{v}\right)\left(\partial v+\partial\xi v\right)\,.
\ee
Note that this expression for the Euclideanised worldsheet action differs from that of~\cite{MO2} by $\partial\leftrightarrow\bar{\partial}$. This is equivalent to a sign flip of the B-field, thereby leading to opposite chiralities of the WZW currents relative to that in~\cite{MO2}. With this difference, the same derivation as in~\cite{MO2} leads to the partition function~\eqref{eqn:StartingPartFunc} with the aforementioned choice of $U_{n,m}$.

\section{Expansion of the partition function}\label{sec:AppendExpansion}
To rewrite the form~\eqref{ThePartFun} of the thermal ${\text{AdS}}_3$ partition function in a manner that displays the conformal dimensions, we begin by expanding the $|\vartheta_1|^{-2}$ factor. The infinite product representation 
\be
\vartheta_1(v|\tau)=-2q^{1/8}\sin\p v\prod_{p=1}^{\infty}(1-e^{2i\p v}q^p)(1-q^p)(1-e^{-2i\p v}q^p)\,,\label{eqn:vartheta1prodrep}
\ee
where $q=e^{2\p i\tau}$, is useful to perform this expansion. Consider
\begin{multline}
\frac{1}{|\vartheta(v|\tau)|^2}=\frac{(q\bar{q})^{-1/8}}{4\sin(\p v)\sin(\p\bar{v})}\prod_{p=1}^{\infty}\frac{1}{(1-q^p)(1-\bar{q}^p)}\\
\times\frac{1}{(1-e^{2iv\p}q^p)(1-e^{-2iv\p}q^p)(1-e^{-2i\bar{v}\p}\bar{q}^p)(1-e^{2i\bar{v}\p}\bar{q}^p)}\,,\label{eqn:rc1.1}
\end{multline}
where $v=s_1\tau-s_2$ with $s_1,s_2\in(0,1)$. Noting that $0<{\text{Im}}(v)<\tau_2$, each factor in this product can be expanded into a converging geometric series. For example, the prefactor expands to
\be
\begin{aligned}
\frac{1}{4\sin(\p v)\sin(\p\bar{v})}&=\frac{e^{-2\p s_1\tau_2}}{(1-e^{2i\p(s_1\tau_1-s_2)}e^{-2\p s_1\tau_2})(1-e^{-2i\p(s_1\tau_1-s_2)}e^{-2\p s_1\tau_2})}\\
&=e^{-2\p s_1\tau_2}\sum_{P_{+,0},P_{-,0}=0}^\infty e^{2i\p(s_1\tau_1-s_2)(P_{+,0}-P_{-,0})}e^{-2\p s_1\tau_2(P_{+,0}+P_{-,0})}\,.
\end{aligned}
\ee 
Similarly, the subsequent factors may be expanded to obtain
\be
\frac{1}{|\vartheta(s_1\tau-s_2|\tau)|^2}=\sum_{q,\bar{q},N,\bar{N}}e^{2\p is_2(q-\bar{q})+2\p i\tau_1[-s_1(q-\bar{q})+N-\bar{N}]-2\p\tau_2[s_1(1-q-\bar{q})-\frac{1}{4}+N+\bar{N}]}\,,\label{DenominatorExpand}
\ee 
by introducing six summation variables $P^+_{\pm,p},P^-_{\pm,p},P_{\pm,p}\in\{0,1,2,\ldots\}$ for each term in the product labelled by $p\in\{1,2,3,\ldots\}$. For compactness of notation, these infinitely many sums (together with those over $P_{\pm,0}$) are denoted by $\sum_{q,\bar{q},N,\bar{N}}$ where $(q,\bar{q})$ and $(N,\bar{N})$ are defined as
\begin{subequations}\label{eqn:degeneracymatch}
\begin{align}
q&=P_{+,0}+\sum_{p=1}^{\infty}(P^+_{+,p}-P^-_{+,p})\,,&\bar{q}&=P_{-,0}+\sum_{p=1}^{\infty}(P^+_{-,p}-P^-_{-,p})\,,\label{eqn:degeneracymatch:qbarq}\\
N&=\sum_{p=1}^{\infty}p(P_{+,p}+P^+_{+,p}+P^-_{+,p})\,,&\bar{N}&=\sum_{p=1}^{\infty}p(P_{-,p}+P^+_{-,p}+P^-_{-,p})\,.
\end{align}
\end{subequations}
Substituting this expansion~\eqref{DenominatorExpand} into~\eqref{ThePartFun} and integrating over $s_2$ simply introduces a Kronecker delta $\d_{a,q-\bar{q}}$ that collapses the sum over $a\in\mathbb{Z}$. At this stage, the identity
\be
e^{B^2/4A}=\sqrt{\frac{A}{\p}}\int_{-\infty}^\infty\mkern-18muds\;e^{-As^2+Bs}
\ee
with $A=\frac{4\p\t_2}{k-2}$ and $B=4\p is_1\t_2$ may be used to introduce an integral over the variable $s$. (This variable shall label the quadratic Casimir of continuous series representations.) This linearises the exponent in $s_1$ facilitating its integration:
\begin{multline}
    2\p\t_2\int_0^1ds_1\;e^{-2\p\t_2s_1\left[-2is+kw_1-i(q-\bar{q})\m-\frac{2\p im'}{\b}-\frac{\m k\b n}{2\p}+1-q-\bar{q}\right]}\\    =\tfrac{e^{-2\p\t_2\left[-2is+kw_1-i(q-\bar{q})\m-\frac{2\p im'}{\b}-\frac{\m k\b n}{2\p}+1-q-\bar{q}\right]}}{2is-kw_1+i(q-\bar{q})\m+\frac{2\p im'}{\b}+\frac{\m k\b n}{2\p}-1+q+\bar{q}}-\tfrac{1}{2is-kw_1+i(q-\bar{q})\m+\frac{2\p im'}{\b}+\frac{\m k\b n}{2\p}-1+q+\bar{q}}\,.\label{s1Integral}
\end{multline}
Using this result in the expansion of the partition function and then performing the change of variables
\be
\tilde{w}_1=w_1+1\,,\qquad\tilde{N}=N-q\,,\qquad\tilde{\bar{N}}=\bar{N}-\bar{q}\,,\qquad\tilde{s}=s-\frac{i(k-2)}{2}
\ee
only in the first term arising from~\eqref{s1Integral}, results in the expression
\be
\begin{aligned}
Z(\b,\m;\t,\bar{\t})&=\sum_{m^\prime,n,w_1\in\mathbb{Z}}\sum_{q',\bar{q}',N,\bar{N}}\Biggl\{\int_{-\infty-\frac{i(k-2)}{2}}^{\infty-\frac{i(k-2)}{2}}\frac{ds}{\p}\sum_{P_{\pm,0}=-\infty}^0\tfrac{1}{2is-kw_1+\left(i(q-\bar{q})\m+\frac{2\p i}{\b}m^\prime+\frac{\m k\b n}{2\p}\right)+1+q+\bar{q}}\\
&-\int_{-\infty}^\infty\frac{ds}{\p}\sum_{P_{\pm,0}=0}^\infty\tfrac{1}{2is-kw_1+\left(i(q-\bar{q})\m+\frac{2\p i}{\b}m^\prime+\frac{\m k\b n}{2\p}\right)-1+q+\bar{q}}\Biggr\}\\
&\times e^{2\p i\t_1\left[w_1(q-\bar{q})+nm^\prime+N-\bar{N}\right]}\\
&\times e^{-2\p\t_2\left[\frac{2s^2}{k-2}-\frac{1}{4}-w_1\left(i(q-\bar{q})\m+\frac{2\p i}{\b}m^\prime+\frac{\m k\b n}{2\p}\right)+\frac{k}{2}w_1^2+\frac{k}{2}\left(\frac{\b n}{2\p}\right)^2(\m^2+1)+i\m m^\prime n+\frac{i\b n(q-\bar{q})}{2\p}(1+\m^2)+N+\bar{N}\right]}\,,
\end{aligned}
\ee
for the partition function. The fact that the sum over $P_{\pm,0}$ in the first line runs over non-positive integers is a consequence of the aforementioned change of variables. Finally, adding and subtracting
\be
\int_{-\infty}^\infty\frac{ds}{\p}\sum_{P_{\pm,0}=-\infty}^0\frac{1}{2is-kw_1+\left(i(q-\bar{q})\m+\frac{2\p i}{\b}m^\prime+\frac{\m k\b n}{2\p}\right)+1+q+\bar{q}}
\ee
in the braces results in an integral over a closed contour that collects residues of the poles located at $\frac{1}{2}+is=\frac{k}{2}w_1-\frac{1}{2}\left(i(q-\bar{q})\m+\frac{2\p i}{\b}m^\prime+\frac{\m k\b n}{2\p}\right)-\frac{1}{2}(q+\bar{q})$ provided that they lie inside to contour of $s$-integration i.e., $-\frac{(k-2)}{2}<\mathfrak{I}(s)<0$. The sum over these residues shall count the degeneracies of discrete series representations and the rest of the terms in braces will become the density of states for continuous series representations. This is precisely the claimed expansion~\eqref{MOG2PartFuncChiralityFlip} that allows us to read off the conformal dimensions.





\begin{thebibliography}{99}

\bibitem{BTZ}
M.~Banados, M.~Henneaux, C.~Teitelboim and J.~Zanelli,
``Geometry of the (2+1) black hole,''
Phys. Rev. D \textbf{48}, 1506-1525 (1993)
[erratum: Phys. Rev. D \textbf{88}, 069902 (2013)]
doi:10.1103/PhysRevD.48.1506
[arXiv:gr-qc/9302012 [gr-qc]].


\bibitem{NS}
M.~Natsuume and Y.~Satoh,
``String theory on three-dimensional black holes,''
Int. J. Mod. Phys. A \textbf{13} (1998), 1229-1262
doi:10.1142/S0217751X98000585
[arXiv:hep-th/9611041 [hep-th]].

\bibitem{Malstrom}
J.~M.~Maldacena and A.~Strominger,
``AdS(3) black holes and a stringy exclusion principle,''
JHEP \textbf{12} (1998), 005
doi:10.1088/1126-6708/1998/12/005
[arXiv:hep-th/9804085 [hep-th]].

\bibitem{MO1}
J.~M.~Maldacena and H.~Ooguri,
``Strings in AdS(3) and SL(2,R) WZW model 1.: The Spectrum,''
J. Math. Phys. \textbf{42} (2001), 2929-2960
doi:10.1063/1.1377273
[arXiv:hep-th/0001053 [hep-th]].
\bibitem{MO2}
J.~M.~Maldacena, H.~Ooguri and J.~Son,
``Strings in AdS(3) and the SL(2,R) WZW model. Part 2. Euclidean black hole,''
J. Math. Phys. \textbf{42} (2001), 2961-2977
doi:10.1063/1.1377039
[arXiv:hep-th/0005183 [hep-th]].

\bibitem{Skenderis}
S.~de Haro, S.~N.~Solodukhin and K.~Skenderis,
``Holographic reconstruction of space-time and renormalization in the AdS / CFT correspondence,''
Commun. Math. Phys. \textbf{217} (2001), 595-622
doi:10.1007/s002200100381
[arXiv:hep-th/0002230 [hep-th]].

\bibitem{Israel}
D.~Israel, C.~Kounnas and M.~P.~Petropoulos,
``Superstrings on NS5 backgrounds, deformed AdS(3) and holography,''
JHEP \textbf{10} (2003), 028
doi:10.1088/1126-6708/2003/10/028
[arXiv:hep-th/0306053 [hep-th]].

\bibitem{Nippanikar}
O.~V.~Nippanikar, A.~Sharma and K.~P.~Yogendran,
``The BTZ black hole spectrum and partition function,''
[arXiv:2112.11253 [hep-th]].

\bibitem{Polchinski}
J. Polchinski, ``String Theory, vol I", Section 8.4


\bibitem{Polchinski2}
J. Polchinski, ``String Theory, vol I", Section 7.3


\bibitem{Buscher}
T.~H.~Buscher,
``Path Integral Derivation of Quantum Duality in Nonlinear Sigma Models,''
Phys. Lett. B \textbf{201} (1988), 466-472
doi:10.1016/0370-2693(88)90602-8

\bibitem{Berkooz:2007fe}
M.~Berkooz, Z.~Komargodski and D.~Reichmann,
``Thermal AdS(3), BTZ and competing winding modes condensation,''
JHEP \textbf{12} (2007), 020
doi:10.1088/1126-6708/2007/12/020
[arXiv:0706.0610 [hep-th]].

\bibitem{sashok}
S.~K.~Ashok and J.~Troost,
``Twisted Strings in Three-dimensional Black Holes,''
[arXiv:2112.08784 [hep-th]].

\bibitem{Esko}
S.~Hemming and E.~Keski-Vakkuri,
``The Spectrum of strings on BTZ black holes and spectral flow in the SL(2,R) WZW model,''
Nucl. Phys. B \textbf{626} (2002), 363-376
doi:10.1016/S0550-3213(02)00021-4
[arXiv:hep-th/0110252 [hep-th]].

\bibitem{RR}
M.~Rangamani and S.~F.~Ross, ``Winding tachyons in BTZ", Phys. Rev. D {\bf 77} (2008) 026010. doi:10.1103/PhysRevD.77.026010 [arXiv:0706.0663 [hep-th]].


\bibitem{McGreevy}
J.~McGreevy,
``Generalized Symmetries in Condensed Matter,''
[arXiv:2204.03045 [cond-mat.str-el]].

\bibitem{Karch}
A.~Karch and B.~Robinson,
``Holographic Black Hole Chemistry,''
JHEP \textbf{12} (2015), 073
doi:10.1007/JHEP12(2015)073
[arXiv:1510.02472 [hep-th]].

\bibitem{DMMV}
R.~Dijkgraaf, J.~M.~Maldacena, G.~W.~Moore and E.~P.~Verlinde,
``A Black hole Farey tail,''
[arXiv:hep-th/0005003 [hep-th]].

\bibitem{MM}
J.~Manschot and G.~W.~Moore,
``A Modern Farey Tail,''
Commun. Num. Theor. Phys. \textbf{4} (2010), 103-159
doi:10.4310/CNTP.2010.v4.n1.a3
[arXiv:0712.0573 [hep-th]].


\end{thebibliography}
\end{document}